\documentclass[aps,prc,twocolumn,superscriptaddress,showpacs,amsmath,amssymb,floatfix]{revtex4}

\usepackage{graphicx}% Include figure files
\usepackage{dcolumn}% Align table columns on decimal point
\usepackage{bm}% bold math
\usepackage{epsfig,amsmath,subfigure}

\def\set{\sigma_{et}}
\def\se{\sigma_{e}}
\def\st{\sigma_{t}}
\def\so{\sigma_{0}}
\def\deg{^\circ}
\def\percent{\% }
\def \percentpunct{\%}
\def\Cerenkov{Cherenkov }
\def\gevm{{\rm GeV/c}^2}

\def\gev{{\rm GeV}}

\def\gevp{{\rm GeV/c}}

\def\gevpsq{{\rm GeV}^2/{\rm c}^2}
\def\carb{${}^{12}\rm{C}$ }
\def\nitr{${}^{15}\rm{N}$ }
\def\nh{${}^{15}\rm{NH}_3$ }
\def\hydr{$\rm{H}$ }
\def\hel{${}^{4}\rm{He}$ }
\def\delt{$\Delta(1232)$ }
\def\npp{N_{\uparrow\uparrow}}
\def\npm{N_{\uparrow\downarrow}}
\def\nmp{N_{\downarrow\uparrow}}
\def\nmm{N_{\downarrow\downarrow}}

\begin{document}
%\preprint{}

\title{Study of $ep\rightarrow ep\pi^{\circ}$ in the $\Delta(1232)$ mass region using polarization asymmetries}

\newcommand*{\uvch}{University of Virginia, Charlottesville, VA 22903, USA}
\newcommand*{\latech}{Center for Applied Physics Studies, Louisiana
Tech University, Ruston, LA 71272, USA}
\newcommand*{\cnuva}{Christopher Newport University, Newport News, VA 23606, USA}
\newcommand*{\jlab}{Thomas Jefferson National Accelerator Facility, 12000 Jefferson Avenue, Newport News, VA 23606, USA}
\newcommand*{\yerevan}{Yerevan Physics Institute, 375036 Yerevan, Armenia}
\newcommand*{\asuaz}{Arizona State University, Tempe, AZ 85287, USA}
\newcommand*{\cmupa}{Carnegie Mellon University,  Pittsburgh, PA 15213, USA}
\newcommand*{\cuawdc}{Catholic University of America, Washington D.C., 20064, USA}
\newcommand*{\cwm}{College of William and Mary, Williamsburg, VA 23187, USA}
\newcommand*{\duke}{Duke University, Physics Bldg. TUNL, Durham, NC 27706, USA}
\newcommand*{\edinburgh}{Edinburgh University, Edinburgh EH9 3JZ, United Kingdom}
\newcommand*{\fiu}{Florida International University, Miami, FL 33199, USA}
\newcommand*{\fsu}{Florida State University,Tallahassee, FL 32306, USA}
\newcommand*{\gwudc}{George Washington University, Washington D. C., 20052 USA}
\newcommand*{\frascati}{Istituto Nazionale di Fisica Nucleare, Laboratori Nazionali di Frascati, P.O. 13, 00044 Frascati, Italy}
\newcommand*{\genova}{Istituto Nazionale di Fisica Nucleare, Sezione di Genova
 e Dipartimento di Fisica dell'Universita, 16146 Genova, Italy}
\newcommand*{\itep}{Institute of Theoretical and Experimental Physics, 25 B. Cheremushkinskaya, Moscow, 117259, Russia}
\newcommand*{\ipn}{Institut de Physique Nucleaire d'Orsay, IN2P3, BP 1, 91406 Orsay, France}
\newcommand*{\jmuva}{James Madison University, Department of Physics, Harrisonburg, VA 22807, USA}
\newcommand*{\knukorea}{Kyungpook National University, Taegu 702-701, South Korea}
\newcommand*{\mmit}{M.I.T.-Bates Linear Accelerator, Middleton, MA 01949, USA}
\newcommand*{\nsuva}{Norfolk State University, Norfolk VA 23504, USA}
\newcommand*{\ohio}{Ohio University, Athens, OH 45701, USA}
\newcommand*{\oduva}{Old Dominion University, Norfolk VA 23529, USA}
\newcommand*{\rpi}{Rensselaer Polytechnic Institute, Troy, NY 12180, USA}
\newcommand*{\rubltx}{Rice University, Bonner Lab, Box 1892, Houston, TX 77251, USA}
\newcommand*{\sphn}{CEA Saclay, DAPNIA-SPhN, F91191 Gif-sur-Yvette Cedex, France}
\newcommand*{\ucla}{University of California at Los Angeles, Los Angeles, CA 90095, USA}
\newcommand*{\connecticut}{University of Connecticut, Storrs, CT 06269, USA}
\newcommand*{\uc}{Union College, Schenectady, NY 12308, USA}
\newcommand*{\umma}{University of Massachusetts, Amherst, MA 01003, USA}
\newcommand*{\umosk}{University of Moscow, Moscow, 119899 Russia}
\newcommand*{\unhdurham}{University of New Hampshire, Durham, NH 03824, USA}
\newcommand*{\uppa}{University of Pittsburgh, Pittsburgh, PA 15260, USA}
\newcommand*{\urva}{University of Richmond, Richmond, VA 23173, USA}
\newcommand*{\usc}{University of South Carolina, Columbia, SC 29208, USA}
\newcommand*{\utep}{University of Texas at El Paso, El Paso, Texas 79968, USA}
\newcommand*{\vpsu}{Virginia Polytechnic and State University, Blacksburg, VA 24061, USA}

\affiliation{\rpi}
%\affiliation{\latech}
\affiliation{\asuaz}
\affiliation{\cmupa}
\affiliation{\cuawdc}
\affiliation{\cnuva}
\affiliation{\cwm}
\affiliation{\duke}
\affiliation{\edinburgh}
\affiliation{\fiu}
\affiliation{\fsu}
\affiliation{\gwudc}
\affiliation{\frascati}
\affiliation{\genova}
\affiliation{\itep}
\affiliation{\ipn}
\affiliation{\jmuva}
\affiliation{\jlab}
\affiliation{\knukorea}
\affiliation{\mmit}
\affiliation{\nsuva}
\affiliation{\oduva}
\affiliation{\ohio}
\affiliation{\rubltx}
\affiliation{\sphn}
\affiliation{\uc}
\affiliation{\ucla}
\affiliation{\connecticut}
\affiliation{\umma}
\affiliation{\umosk}
\affiliation{\unhdurham}
\affiliation{\uppa}
\affiliation{\urva}
\affiliation{\usc}
\affiliation{\utep}
\affiliation{\uvch}
\affiliation{\vpsu}
\affiliation{\yerevan}


\author{A.~Biselli}
\email[Contact Author \ ]{biselli@jlab.org}
%\thanks{}
\author{G.S.~Adams}\affiliation{\rpi}
\author{M.J.~Amaryan}\affiliation{\yerevan}
\author{E.~Anciant}\affiliation{\sphn}
\author{M.~Anghinolfi}\affiliation{\genova}
%\author{D.S.~Armstrong}\affiliation{\cwm}
\author{B.~Asavapibhop}\affiliation{\umma}
\author{G.~Asryan}\affiliation{\yerevan}
\author{G.~Audit}\affiliation{\sphn}
\author{T.~Auger}\affiliation{\sphn}
\author{H.~Avakian}\affiliation{\frascati}
% ...

\collaboration{The CLAS Collaboration}
\noaffiliation

\date{\today}

\begin{abstract}
Measurements of the angular distributions of target and double 
spin asymmetries for  the $\Delta^{+}(1232)$ in 
the exclusive channel $\vec{p}(\vec{e},e'p)\pi^{0}$ obtained 
at Jefferson Lab in the $Q^2$ range from 0.5 to 1.5 $\gevpsq$ 
are presented.
Results of the asymmetries are compared with the 
unitary isobar model~\cite{maidref}, dynamical models~\cite{lee2,dmt},
and the effective Lagrangian theory~\cite{davidson}. 
Sensitivity to the different models was observed, 
particularly in relation to the description of background terms 
on which the target 
asymmetry depends significantly.
\end{abstract}

\pacs{13.60.Le, 13.88.+e, 14.20.Gk}
%\keywords{}

\maketitle

%\section{}
\section{Introduction \label{intro}}

The $\Delta(1232)$ resonance has been one of the
most studied objects in nuclear physics. 
As the lowest energy nucleon excitation it dominates 
the low energy cross sections for pion- and electromagnetic-induced reactions,
and is almost completely separated in
excitation energy from the many broad higher mass resonances.
There is extensive theoretical literature attempting
to characterize the electromagnetic excitation of the $\Delta(1232)$.
Examples of some approaches are: effective Lagrangian models
~\cite{lee,lee2,dmt,kamalov-yang,MAID-dynamic,maidref,davidson},
dispersion relations~\cite{aznaurian}, partial-wave-analysis~\cite{arndt},
quark models~\cite{capstick,simula}, QCD sum-rule models~\cite{balyaev}, or
the generalized parton distribution (GPD) approach~\cite{stoler,frank},
and perturbative QCD with QCD sum-rules~\cite{carlson}.
In recent years there has been considerable experimental activity 
using polarized real photons at LEGS~\cite{legs} and 
Mainz~\cite{mainz}, unpolarized electrons
at Bonn~\cite{bonn} and Jefferson Lab (JLab)~\cite{frolov,joo-smith}, 
polarized electrons
at Mainz~\cite{bartch} and JLab~\cite{joo},and polarized electrons
with recoil polarization at Mainz~\cite{pospischil_delta} and  
Bates~\cite{bates}, which
have focused on constraining our understanding of the 
electromagnetic structure of the $\Delta(1232)$. 

It has long been realized that the proper extraction of resonance 
information from experimental data requires an understanding of 
non-resonant contributions in the vicinity of the resonance pole.
Some of the previously-mentioned theoretical approaches have been developed to 
obtain a more realistic description of the full pion production
amplitude and  in particular the determination of the resonance
contributions. It was  found that certain polarization observables 
e.g. single spin asymmetries, where the polarization of only 
one particle is determined, are sensitive to interferences between 
resonant and non-resonant contributions, while double polarization
observables are more constrained by resonant contributions. Both 
contain information not contained in unpolarized cross sections 
alone.

The main aim of this article is to present the results of a measurement
of polarization observables  in single $\pi^{\circ}$ electro-production.
It is expected that these results, together with other data will aid in reaching a better 
understanding of what is the most  appropriate description of the complete
pion production amplitude in the region of the $\Delta(1232)$. 

Among the theoretical approaches which have appeared during the 
past several years with the aim of extracting resonance amplitudes
from existing data is the aforementioned effective Lagrangian model
~\cite{maidref} (MAID) and ~\cite{davidson} (DM), in which the degrees of freedom
are baryon and meson currents. 
These models include pion scattering effects by using 
the K-matrix method to unitarize the amplitude. 
The differences between MAID and DM arise mainly from some rather
significant differences in their starting effective Lagrangians.
In particular, MAID uses a mixture of pseudo-scalar and pseudo-vector
for the $\pi NN$ coupling, while DM uses the standard pseudo-vector
coupling. MAID includes some higher resonances and hence has more freedom in 
fitting the data.

A major controversy which has developed is
that the resonance amplitude calculated in the framework of the
quark model~\cite{capstick} is significantly smaller than that
extracted from effective Lagrangian models. Such a significant difference
($\sim$30\%) for the presumably best understood resonance points
to a very serious shortcoming for the quark model. However, it has
been pointed out by the authors of ref.~\cite{capstick} that the 
quark models - so far - are not able to take into account the coupling of 
the quarks to the pion cloud, and if this were rectified, one would
expect better agreement with the amplitudes extracted from effective
Lagrangian models.

With this in mind, an elaboration~\cite{lee2} of the effective Lagrangian 
model, the {\em dynamic model} (SL), was developed in which the primary 
resonant and non-resonant interactions involving the pion cloud are 
treated in a consistent coupled channel approach to all orders. 
This was followed by analogous {\em dynamic} 
formulations~\cite{MAID-dynamic} (DMT).
The SL model obtains the unitary amplitudes by solving dynamical 
$\pi N$ scattering equations. Thus the pion cloud effects on the 
extracted `dressed' $N-\Delta$ can be identified and
an interpretation of the resulting `bare' parameters in terms 
of constituent quark model calculations has been established. 
The DMT model uses a chiral Lagrangian which includes the pion 
re-scattering in a coupled channel $t-matrix$ approach.

The net result yields a {\em bare} $\Delta(1232)$ resonance 
amplitudes, stripped
of its coupling with non resonant channels {\em dressed} $\Delta(1232)$, 
which is smaller
than obtained in the more traditional effective Lagrangian
formulations, and in better agreement with that obtained obtained with
the  quark model. The coupling to all orders is also effected in
the dispersion relation calculation~\cite{aznaurian}, and again it is
found that the {\em bare} $\Delta(1232)$ agrees better with that of the quark model.
The most important constraints for these models have been the high quality
non-polarized cross sections which have appeared in recent 
years~\cite{frolov, joo-smith}.

The analysis of JLab unpolarized  cross section data~\cite{joo-smith, frolov}
using these various theoretical formalisms yield very different extracted
non-leading amplitudes $Re(E_{1+}/M_{1+})$ and $Re(S_{1+}/M_{1+})$,
depending on the model used. This is especially true with increasing
momentum transfer, i.e. for $Q^2$ in the multi GeV$^2$/c$^2$, where
the relative contribution of the non-resonant amplitudes become
more important relative to the resonant amplitudes. Thus, in order to
obtain confident estimates of the resonant amplitudes one needs to
determine which formulation best accounts for the overall body 
of the world's data.

In addition to the  non-polarized cross sections, these theoretical 
formulations can predict interference cross sections which can only be 
accessed by polarization variables. Of note are the enhanced sensitivities
to interferences between resonant and non-resonant amplitudes. 
Such interferences can offer strong  constraints on models for extracting
the interplay between resonant and non-resonant amplitudes.  
For example, in the case of  the Mainz~\cite{bartch} single electron 
asymmetry data $Q^2$ = 0.2 GeV$^2$/c$^2$, the predictions of  some of 
the above theoretical formulations~\cite{lee2,MAID-dynamic,maidref} 
differ significantly, and none give fully
satisfactory agreements with the data.  The authors speculated
that the treatments of the non-resonant backgrounds may be the cause,
though no quantitative comparisons between the different predictions and
experiment were made. The JLab data~\cite{joo} obtained  at  higher 
$Q^2$ = 0.4 and .65  GeV$^2$/c$^2$ were also compared with
the results of the same theory gave equally divergent results.

In the case of the Mainz ~\cite{pospischil_delta} and
Bates ~\cite{bates} recoil polarization
experiments at $Q^2 \sim 0.1$  GeV$^2$/c$^2$, comparisons were 
made with one of the models (MAID) to extract the $\Delta(1232)$
 quadrupole amplitude
$Re(S_{1+}/M_{1+})$. However, since the different models are
shown to yield different results for  non-leading amplitudes
when compared to other data, it would seem that one would need
better confidence in the theoretical basis. 
 
With this background in mind, the present report provides
independent double polarization data, which will be useful
in testing the models, especially at previously
unexplored higher $Q^2$ (0.5 to 1.5 GeV$^2$/c$^2$),
where  new physics may be opening up, and background effects are 
becoming relatively more important.
The reaction studied in the presently reported experiment is 
$\vec{e}+\vec{p}\to e^\prime + p + \pi^0$,
where the scattered electron and emitted proton were
observed in coincidence, and the $\pi^0$ was identified by the
missing mass technique. Although the feasibility of exclusive coincidence
experiments involving target and beam double polarization was demonstrated in
the reaction $\vec{e}+\vec{p}\to e^\prime + n + \pi^+$ in 
ref.~\cite{raffaella}, this is the first time such experiments are
carried out in which the $Q^2$ behavior of the target and double spin 
asymmetries for a specific resonance are explored in the  $GeV$ range 
of momentum transfer.
We expect these 
unique polarization observables to give significant constraints
for improving theories of the $\Delta(1232)$ electro-production process. 

In addition, quantitative comparisons are made to the predictions
of the four theoretical approaches, MAID~\cite{maidref}, SL~\cite{lee2},
DMT~\cite{dmt} and DM~\cite{davidson},

\section{Formalism}
In this experiment single mesons are produced by a polarized electron beam
 incident on a polarized proton target polarized parallel or antiparallel
to the electron beam direction, as schematically shown in 
Figure~\ref{elecprod}. The incident polarized electron is given by the
 4-vector $p_e=(\vec{p}_e,E_i)$, the outgoing electron is emitted with 
 angles $\phi_{e}$, $\theta_{e}$ and 4-vector 
$p_e'=(\vec{p}_e\,',E_{f})$, the virtual photon is characterized by
 $q=(\vec{q},\omega)$ where $\vec{q}=\vec{p}_e-\vec{p}_e\,'$ and $\omega=E_i-E_f$ and
 the nucleon initial and final states are given by $p_p=(0,M)$ and
 $p_p'=(\vec{p}_p,E_p)$, respectively. In terms of these variables
 the cross section can be written as: 
\begin{figure}[here,top]
\includegraphics[width=3.5in]{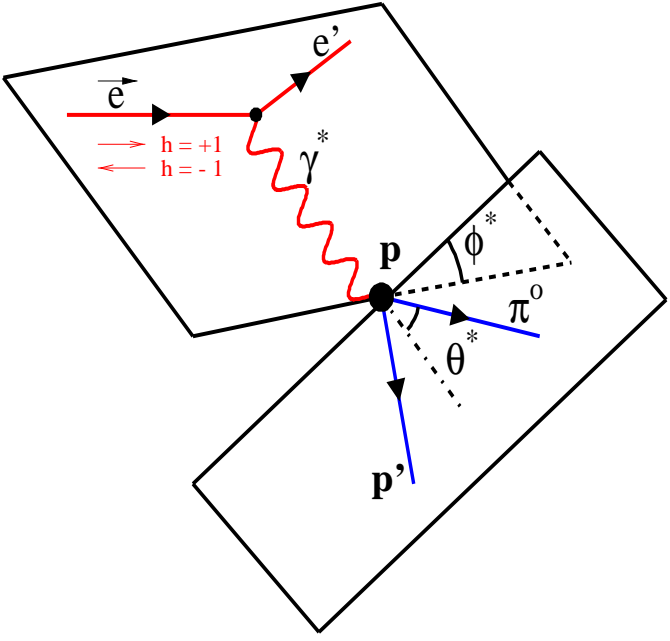}
\caption{(color) Schematic diagram of $\pi$-nucleon electro-production.
 $\vec{e}$ represents the incident polarized electron, $e'$ is the outgoing electron,
 $\gamma^{*}$ is the virtual photon and $p$ and $p'$ are the nucleon in the initial
 and final state, respectively.}

\label{elecprod}
\end{figure}
\begin{equation}
\frac {d\sigma}{d E_f d\Omega_e d\Omega^*}=\Gamma\frac{d\sigma}{d\Omega ^*},
\end{equation} 
where 
$d\Omega_e=\sin \theta_e d\theta_e d\phi_e $ is the electron solid angle, $d\Omega^*=\sin \theta^* d\theta^* d\phi^*$ is the solid angle of the meson in the center of mass, 
\begin{equation}
\Gamma=\frac{\alpha}{2\pi^2}\frac{E_f}{E_i}\frac{k^{lab}_{\gamma}}{Q^2}\frac{1}{1-\epsilon}
\end{equation} 
is the virtual photon flux,
\begin{equation}
\epsilon=(1+2\frac{{\mid\vec{q}\mid}^2}{ Q^2}\tan^2\frac{\theta_e}{2})^{-1}
\end{equation}
represents the degree of polarization of the virtual photon, 
\begin{equation}
k^{lab}_{\gamma}=\frac{W^2-M^2}{2M}
\end{equation}
denotes the `photon equivalent energy' necessary for a real photon 
to excite a hadronic system with center-of-mass (c.m.) energy 
$W=|p_e+p_p-p_e'|$, $Q^2=-q^2=-(\omega^2-\vec{q}^{\ 2})$ is the momentum transfer and $\alpha$ is the fine structure constant.
The differential cross section for pion production by a virtual 
photon $d\sigma / d\Omega^{*}$ can be written as a sum of four terms 
as follows:
\begin{equation}
\label{forsigma2}
\frac{d\sigma}{d\Omega^{*}}= \frac{|\vec{k}|}{k_{\gamma}^{c.m.}}\biggl\{\frac{d\sigma_0}{d\Omega^{*}}+h\frac{d\sigma_e}{d\Omega^{*}}+P\frac{d\sigma_t}{d\Omega^{*}}-hP\frac{d\sigma_{et}}{d\Omega^{*}}\biggr\},
\end{equation}
where $\vec{k}$ is the momentum of the pion, $h$ is the electron helicity and $P$ is the target proton polarization. 
The first term $d\sigma_0 / d\Omega^{*}$ represents the unpolarized cross section, 
while the remaining terms $d\sigma_e / d\Omega^{*}$, $d\sigma_t / d\Omega^{*}$, and $d\sigma_{et} / d\Omega^{*}$ 
arise when beam, target, or both beam and target are polarized, respectively. Here 
\begin{equation}
k_{\gamma}^{c.m.}=\frac{M}{W}k_{\gamma}^{lab}
\end{equation}
is the `real photon equivalent energy' in the c.m. frame.
These cross sections can be written in terms of response functions $R$ using the formalism of reference~\cite{tiator} as:
\begin{widetext}
\begin{equation}
\label{forsigma}
\begin{split}
\frac{d\sigma_0}{d\Omega^{*}}
        =&R_{T}^{0} + \epsilon_{L} R_{L}^{0} +\sqrt{2\epsilon_{L}(1+\epsilon)} R_{TL}^{0} \cos\phi^{*} + \epsilon R_{TT}^{0} \cos 2\phi^{*}\\
\frac{d\sigma_e}{d\Omega^{*}}
        =&\sqrt{2\epsilon_{L}(1-\epsilon)} R_{TL'}^{0} \sin\phi^{*}\\
\frac{d\sigma_t}{d\Omega^{*}}
        =&\sin\theta_{\gamma}\cos\phi^*[\sqrt{2\epsilon_{L}(1+\epsilon)} R_{TL}^{x} \sin\phi^{*} + \epsilon R_{TT}^{x} \sin 2\phi^{*}]
        +\sin\theta_{\gamma}\sin\phi^*[R_{TL}^{y} + \epsilon_{L} R_{L}^{y} +  \\
        +& \sqrt{2\epsilon_{L}(1+\epsilon)} R_{TL}^{y} \cos\phi^{*}+ \epsilon R_{TT}^{y} \cos 2\phi^{*}] +\cos\theta_{\gamma}[\sqrt{2\epsilon_{L}(1+\epsilon)} R_{TL}^{z} \sin\phi^{*} + \epsilon R_{TT}^{z} \sin 2\phi^{*}]\\
\frac{d\sigma_{et}}{d\Omega^{*}}
        =&-\sin\theta_{\gamma}[\sqrt{2\epsilon_{L}(1-\epsilon)} R_{TL'}^{x} {\cos\phi^{*}}^{2} + \sqrt{1-\epsilon^{2}} R_{TT'}^{x}\cos\phi^*]
        +\sin\theta_{\gamma}\sqrt{2\epsilon_{L}(1-\epsilon)} R_{TL'}^{y} {\sin\phi^{*}}^{2}\\
        -&\cos\theta_{\gamma} [\sqrt{2\epsilon_{L}(1-\epsilon)} R_{TL'}^{z} \cos\phi^{*} + \sqrt{1-\epsilon^{2}} R_{TT'}^{z}],\\
\end{split}
\end{equation}
\end{widetext}
where
\begin{equation}
\epsilon_L=\frac{Q^2}{\omega^2}\epsilon
\end{equation}
is the frame-dependent longitudinal polarization the virtual photon. The $\theta_{\gamma}$ is the angle between
the directions of the target polarization and  virtual photon.

The asymmetries are then defined as follows:
\begin{equation}
\begin{split}
A_e=&\frac{\se}{\so}\\
A_t=&\frac{\st}{\so}\\
A_{et}=&\frac{\set}{\so},\\
\end{split}
\end{equation}
where  $\so \equiv d\sigma_0 / d\Omega^{*}$, $\se \equiv d\sigma_e / d\Omega^{*}$, $\st \equiv d\sigma_t / d\Omega^{*}$, and $\set \equiv d\sigma_{et} / d\Omega^{*}$.

\section{Experimental setup}
\label{expsetup}

The experiment was carried out from September to December 1998 
using the CEBAF Large Acceptance
Spectrometer (CLAS) at JLAB, using a polarized electron beam of energy 
$E=2.565\ \gev$ at an average  beam current of about 2 nA. 
Pairs of complementary helicity states were created pseudo-randomly by a pockel cell producing 
circularly polarized laser light used to generate polarized electrons from a strained 
GaAs photocathode \cite{sinclair}.  Each pair of complementary helicity states had a duration of 2 sec. 
Helicity-correlated systematic uncertainties are reduced by  
selecting the first helicity of the pair pseudo-randomly.
The average polarization of the beam for the entire data set, measured with a M{\o}ller 
polarimeter, was $P_e=0.71 \pm 0.01$.
The beam was rastered in a spiral pattern of 1-1.2 cm diameter over the surface of the target to avoid destroying 
the target polarization.

The electrons impinged on a solid ammonia ($\rm{NH}_3$) target of thickness
530 $\rm{mg}/\rm{cm}^2$, in which the free protons were longitudinally
polarized. The target polarization was changed every 2-3 weeks.
{\em Dynamic nuclear polarization} (DNP)\cite{DNP,DNP2} was used 
to polarize this target using a 5 T uniform holding-field generated
by a super-conducting Helmholtz-like coil  placed axially around the target.
This coil limited  the available scattering angles 
to less than $45 \deg$ and between $70 \deg$ and $110 \deg$.
A more complete description of the target and polarization technique may be 
found in Ref.~\cite{keith}. Typically, the polarizations achieved for 
positive and negative
polarizations were about 39\percent and 55\percentpunct, respectively.
The effective instantaneous luminosity for the polarized Hydrogen was
about $6.6 \times 10^{32}\ \rm{cm}^{-2}\rm{s}^{-1}$.

Scattered electrons and recoiled protons were detected in the
CLAS, which is described in detail in Ref.~\cite{clas2}.
An event was triggered when a coincidence between the threshold \Cerenkov 
counter (CC) and the electromagnetic calorimeter 
(EC)  was detected.
A typical \Cerenkov signal 
consisted of 6 to 12  photo-electrons (PE),
with an average of about 10.
The trigger threshold was set at 0.5 PE.
Electron candidates were identified by
a combination of time-of-flight (TOF) scintillators,  
CC, and EC. The TOF scintillators completely 
surround the drift chambers whereas the EC and the CC subtend 
angles less than $45\deg$ with respect to the beam line.
The momenta of the detected particles were determined by fitting
their measured trajectories in the toroidal field, which curves the tracks
in the $\theta$ direction but leaves them nearly unaffected in the $\phi$
direction. The trajectories are determined by 3 sets of drift
chambers (DC), the inner most having 10 layers and 
the other two having each 12 layers of drift cells. 
   
\section{Data Reduction and Analysis}

\begin{figure*}[here,top]
\centering
\mbox{\subfigure{\epsfig{figure=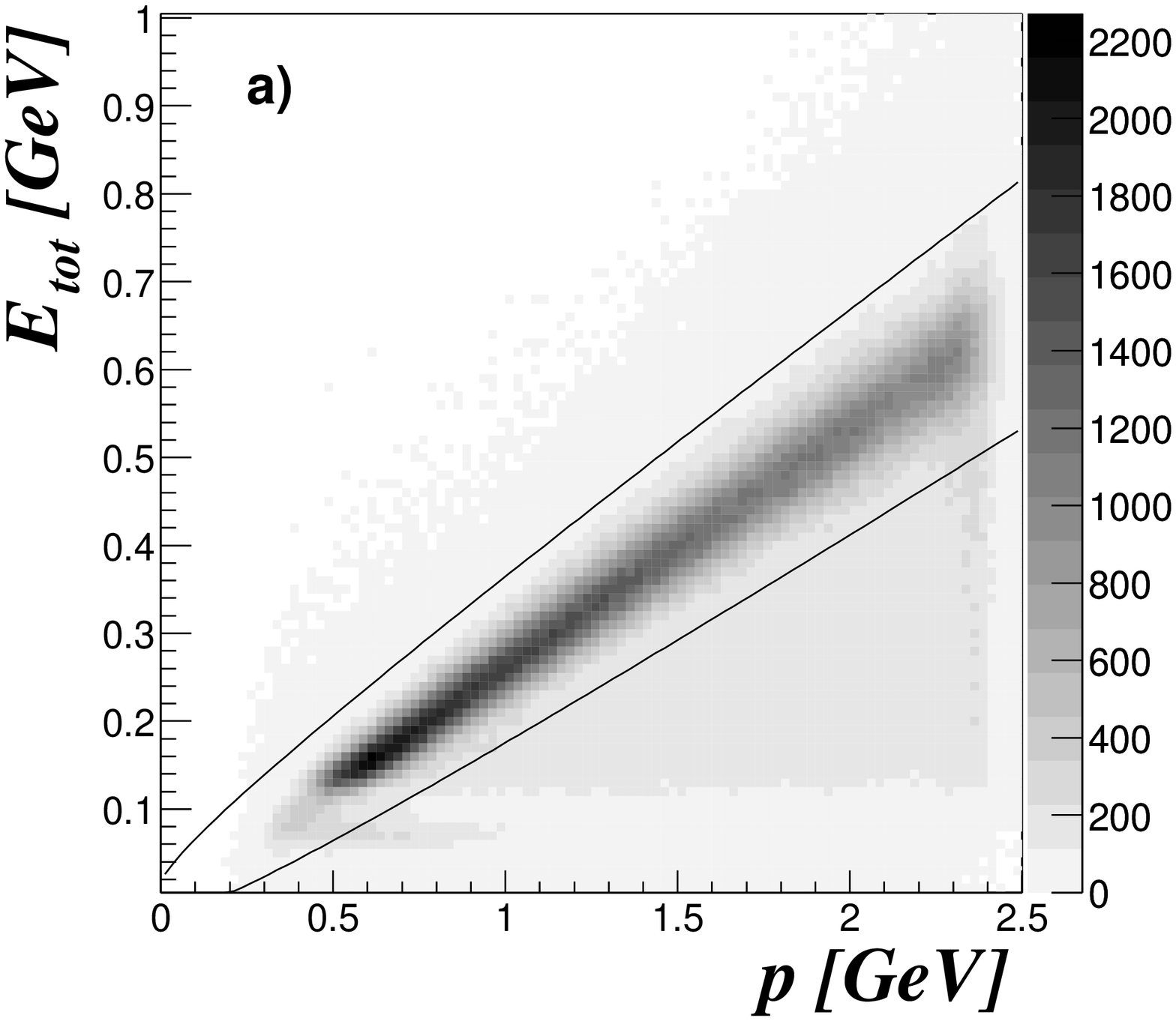,width=2.5in}}\quad
\subfigure{\epsfig{figure=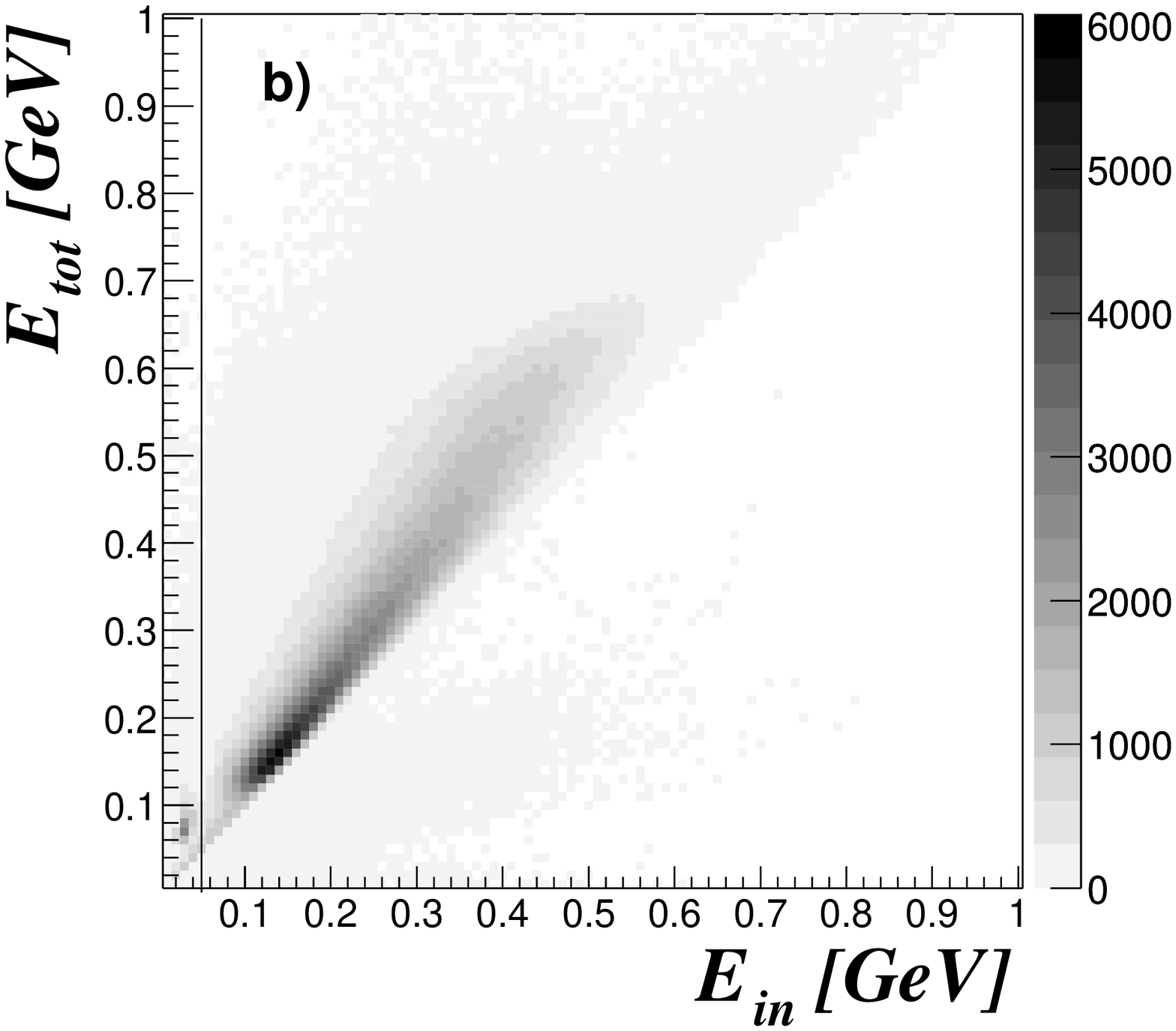,width=2.5in}}}
\caption{{Electron identification. a) $E_{tot}\ vs\ p$. The two lines indicate the cut applied 
to remove events that deviate by more than three sigma from the expected 
behavior. b) $E_{tot}\ vs\ E_{in}$. The line indicates the cut applied to remove events that have 
$E_{in}$ much smaller than $E_{tot}$, which correspond to misidentified pions.}}
\label{ECcut}
\end{figure*}

{\bf Electron identification.} Electron identification 
was improved offline in 
order to remove pions and other sources of contamination.
The EC signal was used to remove events 
in which tracks triggered the CC but did not shower in the EC, such as 
pions which generate secondary electrons.
The energy released by electrons traversing the EC is proportional 
to the momentum $p$ as shown in Figure~\ref{ECcut}(a).
The width of the band is due to EC resolution and the lines indicate
the cut applied to remove background. The EC signal 
is also measured separately for the inner part (15 layers of 
scintillators) and outer part (24 layers).
This allows one to distinguish between an electron, which showers mostly
in the inner part, and minimum ionizing particles, such as $\pi$'s, which lose 
most of their energy in the outer part.
This behavior is evident in Figure~\ref{ECcut}(b) where the high intensity 
region with $E_{in} \sim E_{tot}$ corresponds to electrons, while the small peak 
at low $E_{in}$ corresponds to misidentified pions. The vertical line indicates 
the cut applied to remove misidentified pions.

\begin{figure}[here,top]
\begin{center}
\includegraphics[width=2.5in]{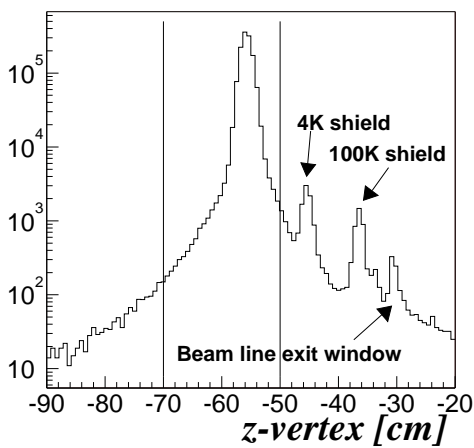}
\end{center}
\caption{The number of events as a function of the vertex z-position 
of the electron where z is along the beamline. 
The lines, which indicate the applied cut, show that the peaks 
from the scattering off the target temperature shields and the beam line exit 
window are completely removed. (Note log vertical scale). The cut 
does not remove the exit and entrance windows from the target cell.}
\label{vertexcut}
\end{figure}

The reconstructed vertex position was used to remove events originating 
from the target temperature shields and the beam line exit window. 
Figure~\ref{vertexcut} shows the cut applied to selected events from 
inside the  target. 

\begin{figure}[here,top]
\begin{center}
\includegraphics[width=2.5in]{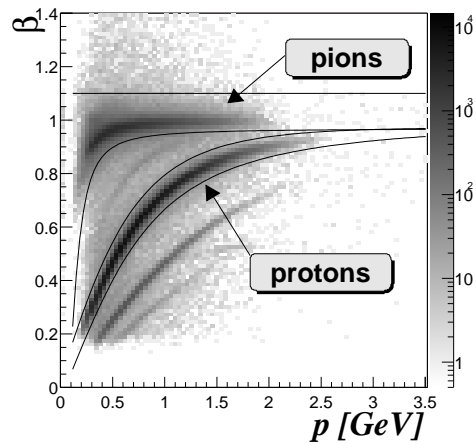}
\end{center}
\caption{$\beta\ vs\ p$ for all positive charge particles. 
The lines show how pions and protons are easily distinguishable.}
\label{protid}
\end{figure}

{\bf Proton identification. }Protons were identified
by determining their momentum and path length using the DC, and their
$\beta=v/c$ using the TOF. Figure~\ref{protid} shows the cut 
applied to select protons, which appear well separated from the pions 
for momenta less than $2\ \gevp$.

\begin{figure*}[here,top]
\centering
\mbox{\subfigure{\epsfig{figure=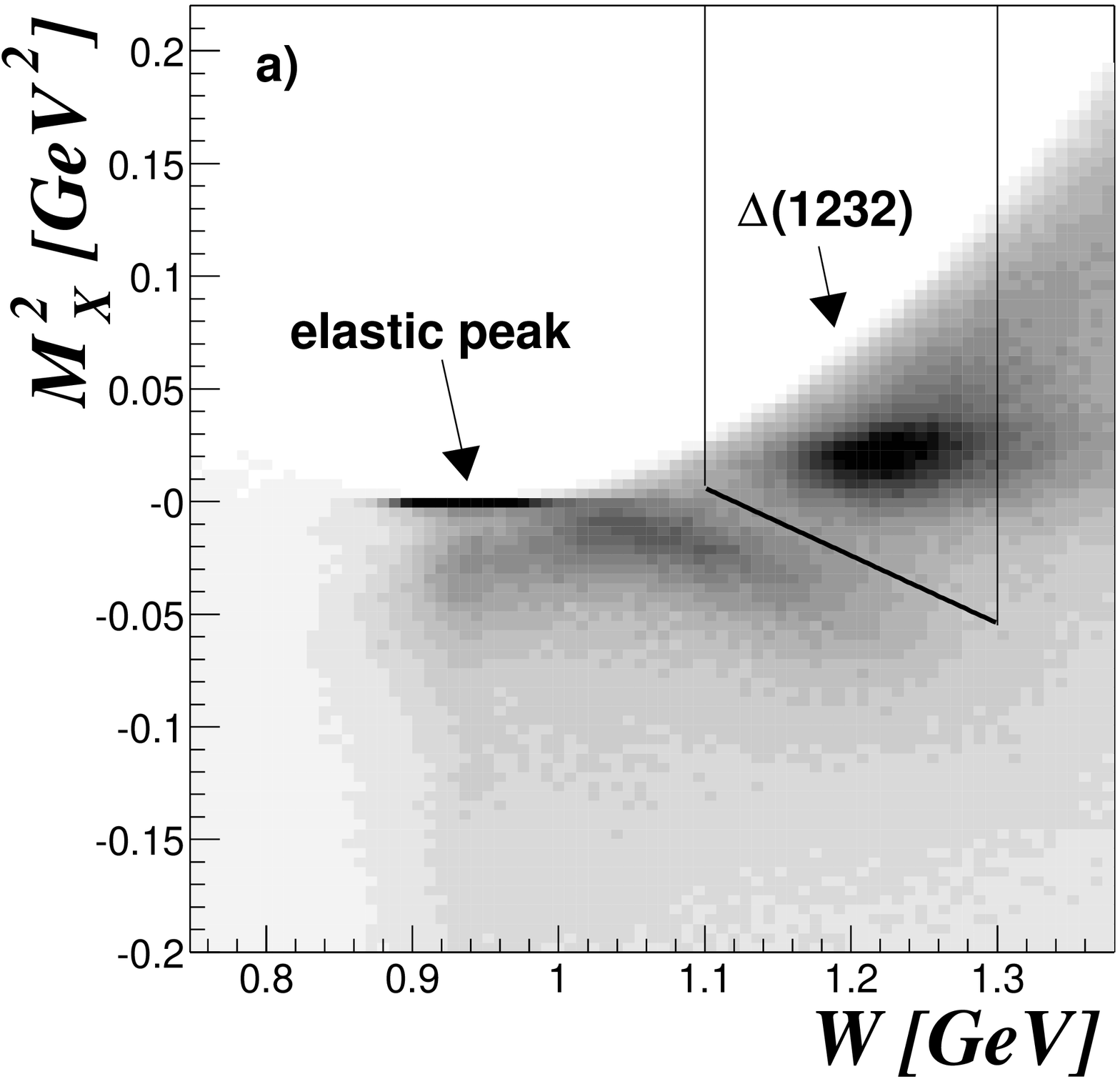,width=2.5in}}\quad
\subfigure{\epsfig{figure=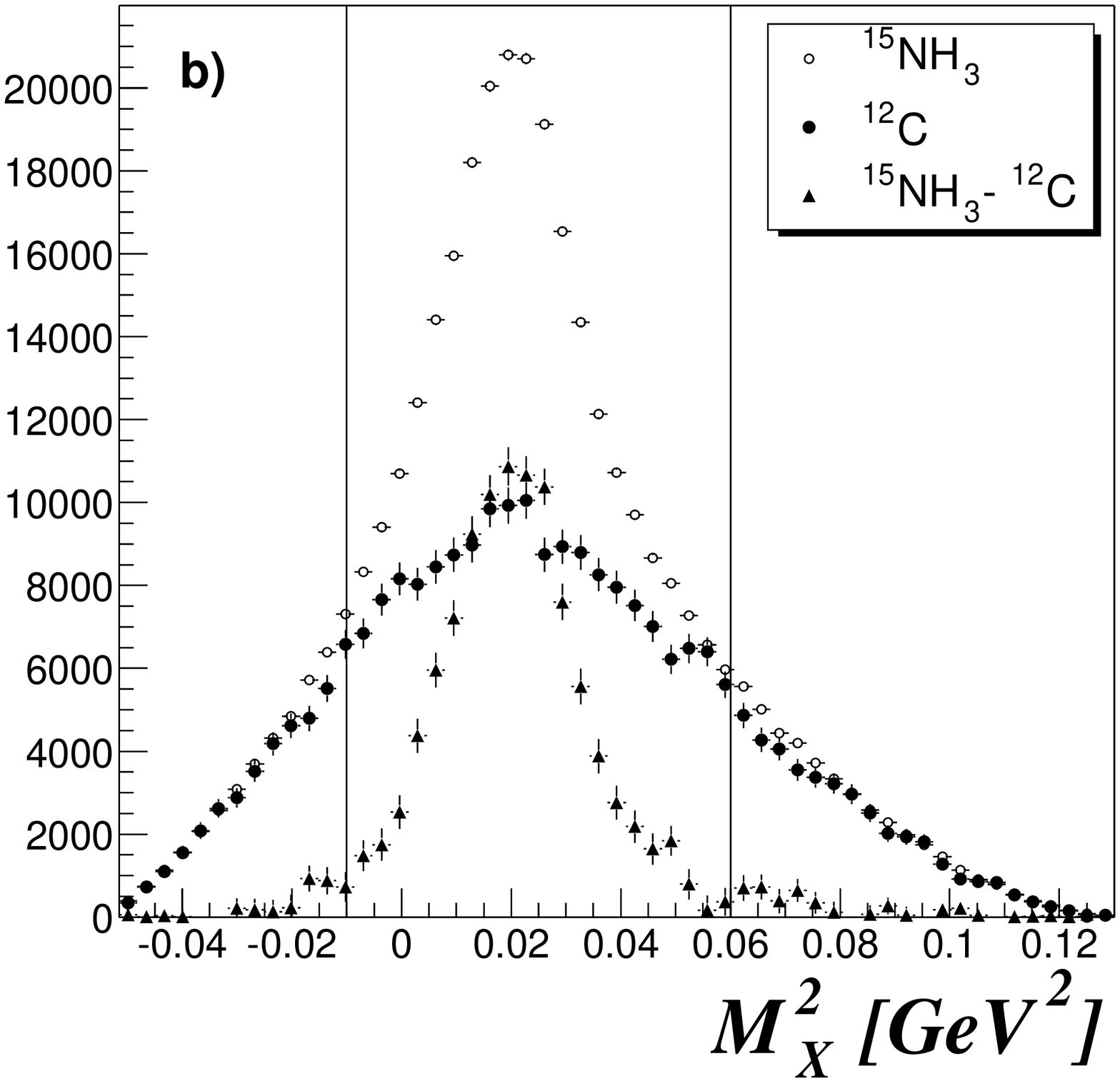,width=2.5in}}}
\caption{{Identification of $p \pi^0$ events. a) $M_X^{2} \ vs \ W$. 
The lines show the two-dimensional cut applied  in order to remove the elastic events 
and quasi-elastic shoulder. b) The plot shows the 
resulting $M_X^{2}$ spectrum after the two-dimensional cut 
(open circles), the \carb data normalized to the \nh target data (full circles) 
and the difference of the two (triangles). The two lines show the 
final cut in $M_X^{2}$ to select pions scattering off hydrogen.}}
\label{mmvsW}
\end{figure*}

{\bf $\pi^0$ channel identification. }In order to select the \delt resonance 
in the decay channel $\Delta^+\rightarrow\pi^{0}p$, cuts on the 
invariant mass $W$ and the square of the missing mass 
$M_X^2=|p_e+p_p-p_e'-p_p'|^2$ were performed. The  
\nh target  intrinsically has a large background due to scattering 
from bound nucleons in \nitr. Many of these events were removed through 
kinematic cuts. An initial two-dimensional cut was applied to select the 
\delt region and to remove the elastic and quasi-elastic events as shown 
in Figure~\ref{mmvsW}(a). The underlying quasi-$\Delta$ events from \nitr, not 
kinematically separable, were removed by a subtraction process by comparing to
data taken with a \carb target. Figure~\ref{mmvsW}(b) shows the missing mass 
spectrum obtained with \nh and \carb targets after the 2-dimensional 
cut and the resulting subtraction. The remaining pion peak due to the 
\hydr is narrower than the \nh peak. A second and much tighter cut on 
$M_X^2$ alone was therefore performed to optimize the selection of 
pions from reactions on free hydrogen in \nh. The two vertical lines 
in Figure~\ref{mmvsW}(b) show the applied cut. 

{\bf Elastic radiative tail. }The elastic radiative 
tail was suppressed by the presence of the target magnetic 
coils that block polar angles between $45\deg$ and $70\deg$. The remaining 
elastic radiative events were removed by means of a cut on the reconstructed 
electron scattering angle ($\theta$)~\cite{angela}. This cut removed 
15\percent of the original data set.

\begin{figure*}[here,top]
\centering
\mbox{\subfigure{\epsfig{figure=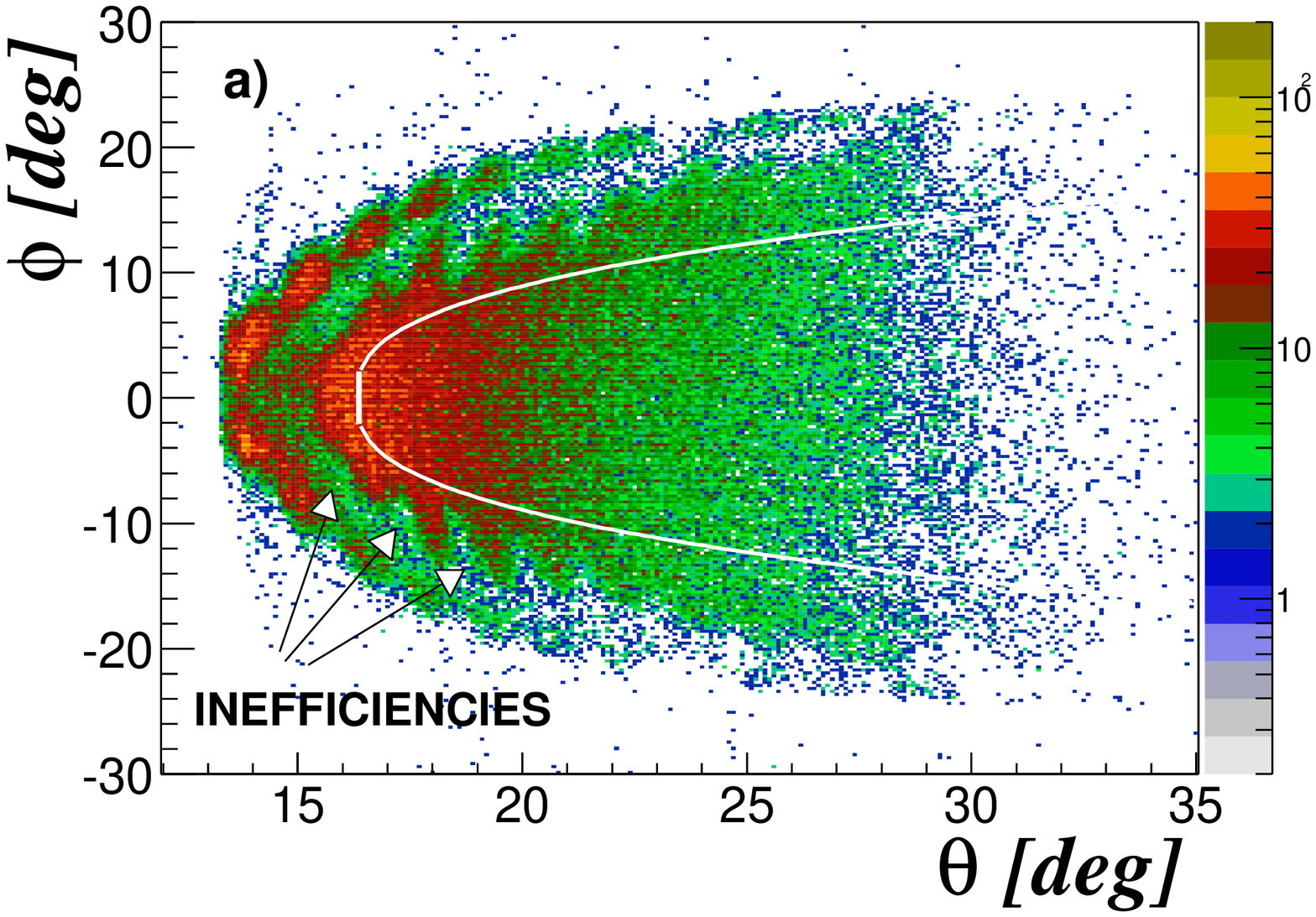,width=2.7in}}\quad
\subfigure{\epsfig{figure=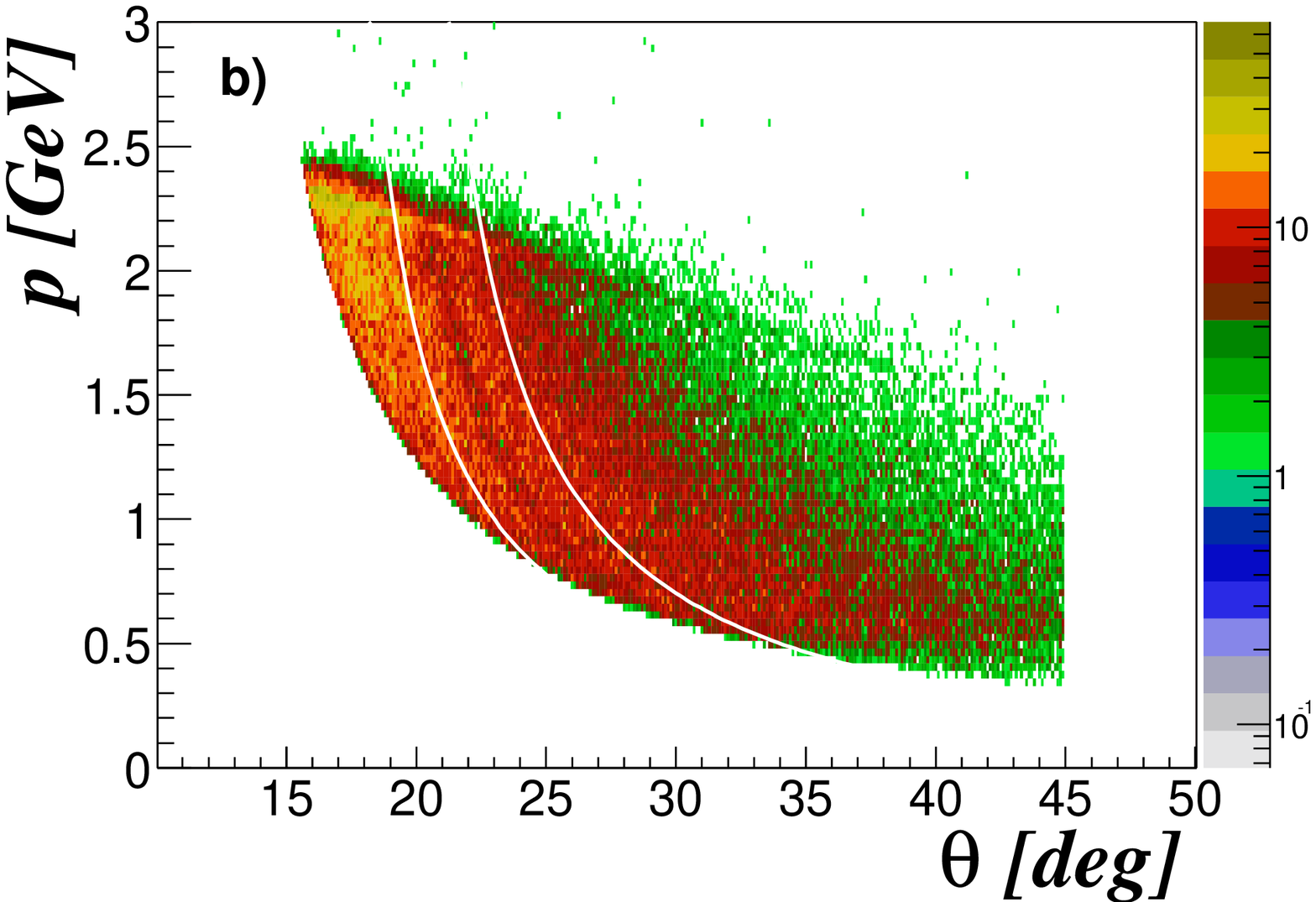,width=2.7in}}}
\caption{{(color) a) $\phi$ vs $\theta$ for electrons in the first CLAS sector for 
a momentum ($p$) bin from $1.9$ to $2.1\ \gevp$. The line indicates the cut 
applied to remove the external fringes and the depletion due to CC 
inefficiencies. b) $p$ vs $\theta$ for electrons  in the third CLAS sector after applying the cut shown in a). 
The region inside the two lines corresponds to an inefficient scintillator.}}
\label{fidcut}
\end{figure*}

\begin{figure*}[here,top]
\centering
\mbox{\subfigure[$0.5\ \gevpsq<Q^2<0.9\ \gevpsq$]{\epsfig{figure=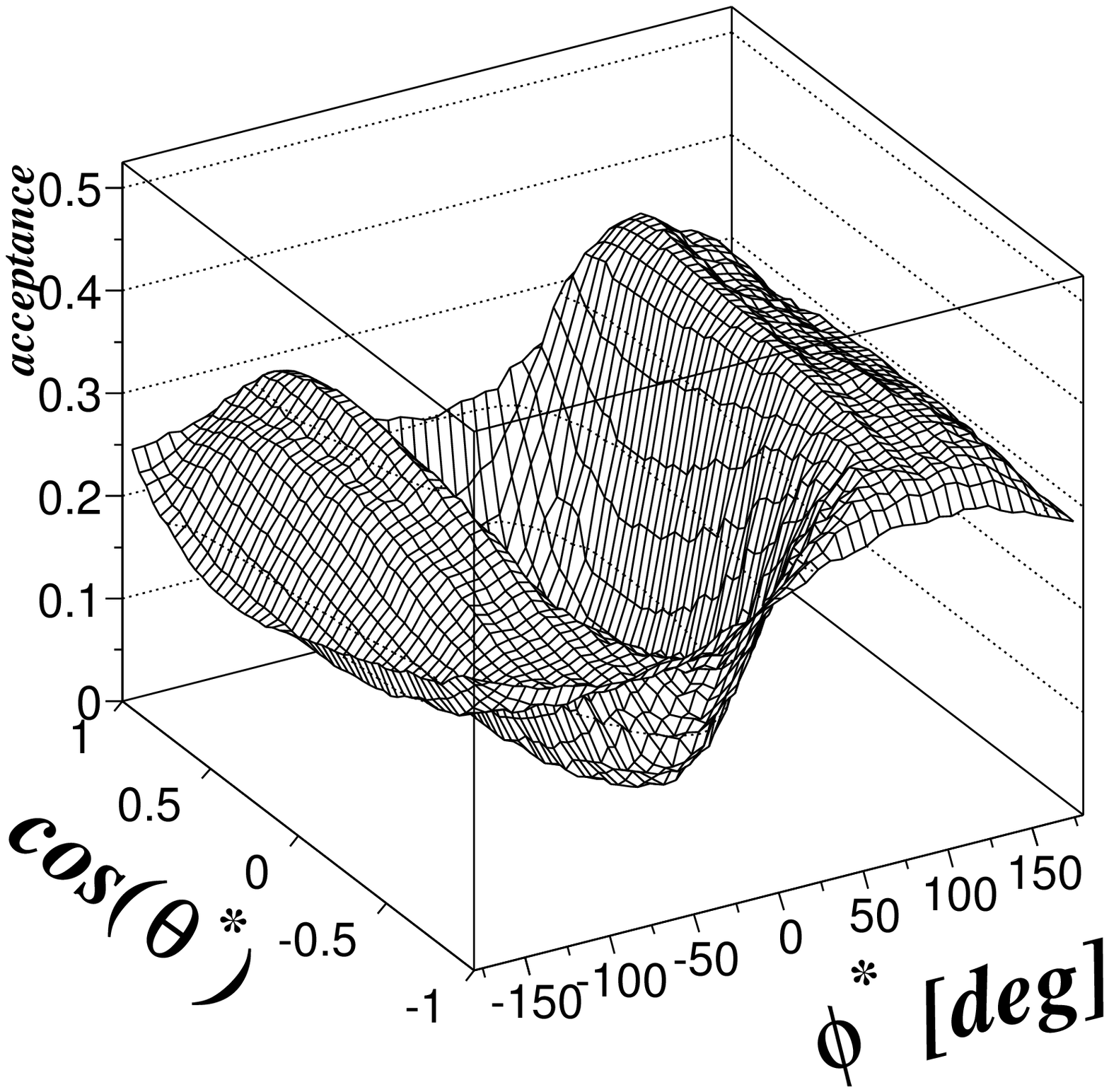,width=2.5in}}\quad
\subfigure[$0.9\ \gevpsq<Q^2<1.5\ \gevpsq$]{\epsfig{figure=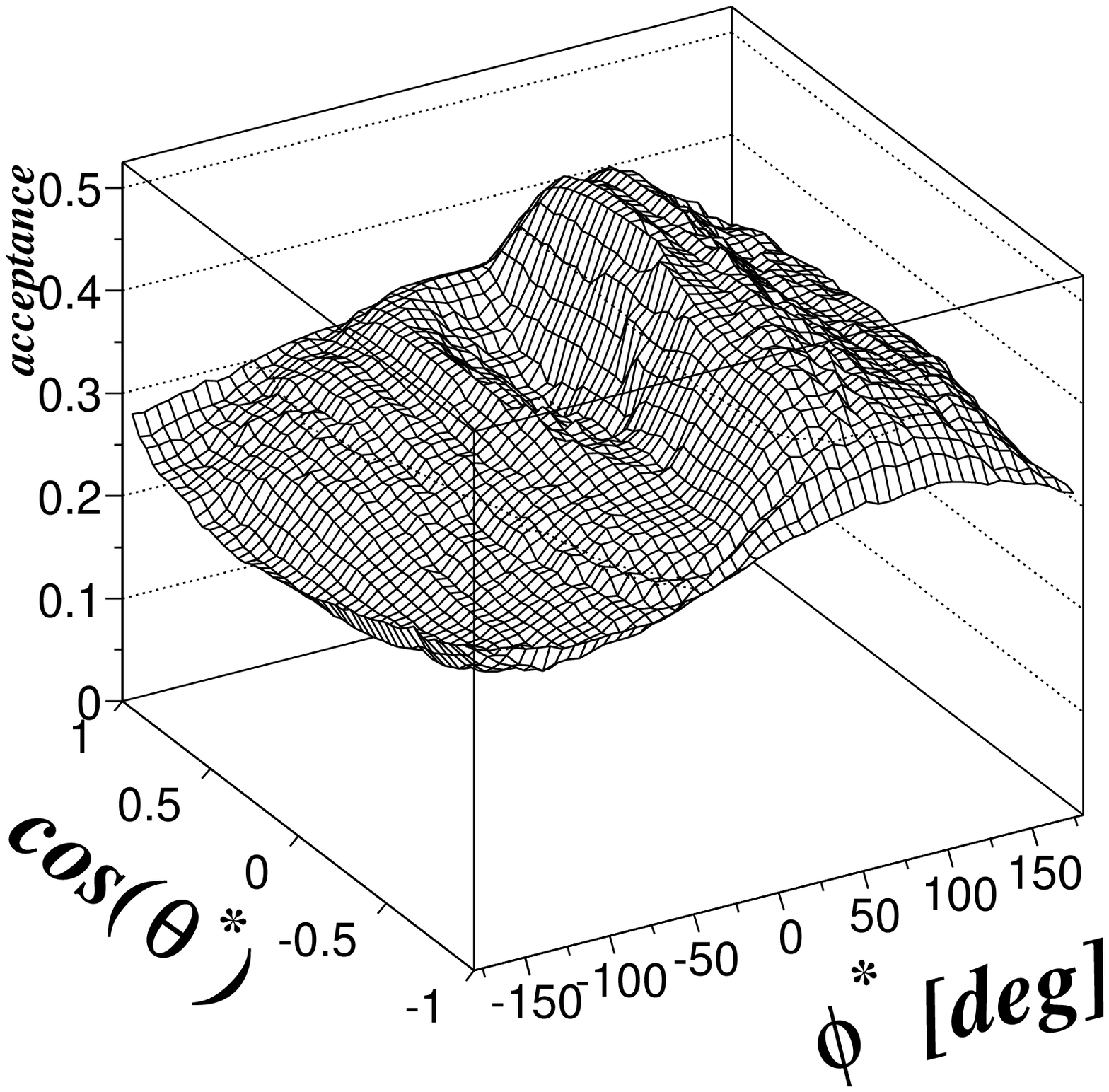,width=2.5in}}}
\caption{{Acceptance calculation for two intervals in $Q^2$ for $1.1\ \gevm<W<1.3\ \gevm$. The lower interval has a region around $\phi^*=0 \deg$ where the acceptance is zero.}}
\label{acc}
\end{figure*}

{\bf Fiducial cuts and acceptance corrections. }The efficiency
can vary by more than an order of magnitude near the boundaries 
of the six azimuthal sectors of CLAS, 
therefore only events in the region where the acceptance is uniform were included. 
Limiting electrons to this fiducial region, gives an elastic scattering
cross section that is consistent with the world's data  to within a few 
percent.
Although the objective of the present analysis is to extract 
asymmetries, a good understanding of the acceptance is necessary.
Calculating the asymmetries involves integrations over ranges in $Q^2$,
$\phi^*$, $\theta^*$ and $W$, and since the acceptance is a function 
of these variables, it does not cancel out when ratios of the integrated
quantities are taken.
Fiducial cuts define a region in $\theta$ and  $\phi$ depending 
on the momentum for both the electron and proton. The area inside 
the line in Figure~\ref{fidcut}(a) is an example of the region selected 
by the fiducial cuts for  electrons detected in the first CLAS sector and 
with momenta between $1.9\ \gevp$ and $2.1\ \gevp$. The cuts not 
only remove data close to the sector boundaries but further remove events from regions 
where scintillators are inefficient  or which have other tracking inefficiencies. 
Figure~\ref{fidcut}(b) displays the effect of a cut to remove an inefficient 
scintillator in the third CLAS sector. The total amount of data removed 
by the fiducial cuts for events with one electron and one proton 
and $W<1.4\ \gevm$ is on the order of $60\%$.
Data were $\phi$-acceptance corrected event by event using an analytical 
calculation based on the assumption that acceptance within 
the fiducial region is 100\percentpunct. Figure~\ref{acc} shows the acceptance 
as a function of $\phi^*$ and $\theta^*$ calculated for two intervals
in $Q^2$ within a $W$ range from $1.1\ \gevm$ to $1.3\ \gevm$.  

{\bf Experimental definition of the asymmetries. }
The experimentally measured number of counts, $N_{ij}$, are grouped according 
to different combinations of beam ($i$) and target ($j$)  polarizations. 
Under the assumption of constant efficiency, these may be written in 
terms of  the cross sections in equations~(\ref{forsigma}) as:
\begin{equation}
\begin{split}
\npp \propto (\so+\so^N+P^{}_e\se+P^{}_e\se^N+P^a_t\st-P^{}_eP^a_t\set)\\
\nmp \propto (\so+\so^N-P^{}_e\se-P^{}_e\se^N+P^a_t\st+P^{}_eP^a_t\set)\\
\npm \propto (\so+\so^N+P^{}_e\se+P^{}_e\se^N-P^b_t\st+P^{}_eP^b_t\set)\\
\nmm \propto (\so+\so^N-P^{}_e\se-P^{}_e\se^N-P^b_t\st-P^{}_eP^b_t\set),
\end{split}
\end{equation}
where $\so^N$ and $\se^N$ are the contributions from the scattering 
from \nitr and the liquid helium coolant, and  $P^a$ and $P^b$ are 
the magnitudes of positive and negative target polarizations, respectively.
The left side of these equations ($N_{ij}$) has been normalized to the 
same total beam charge.  The asymmetries may be written in terms of these quantities as:
\begin{equation}
\begin{split}
\label{forasymdef}
A_t=\frac{\st}{\so}=\frac{1}{P^b_t}\frac{(\npp+\nmp)-(\npm+\nmm)}{(\npp+\nmp)+\alpha(\npm+\nmm)-\beta\so^N}\\
A_{et}=\frac{\set}{\so}=\frac{1}{P^{}_eP^b_t}\frac{-(\npp-\nmp)+(\npm-\nmm)}{(\npp+\nmp)+\alpha(\npm+\nmm)-\beta\so^N},
\end{split}
\end{equation}
where
\begin{equation}
\label{forasymdef2}
\alpha=\frac{P^a_t}{P^b_t}
\end{equation}
and
\begin{equation}
\beta=2(1+\alpha)
\end{equation} 
The extraction of the nuclear background cross section $\so^N$ and 
the constant $\alpha$ are discussed in the next two sections.

{\bf Background subtraction. }
The data have a large background $\so^N$ due to scattering from \nitr 
and the helium cooling bath. Data taken with \carb and \hel targets  were used
to remove this contribution. While the \carb and \nitr targets had  similar 
radiation lengths, they displaced different amounts of helium. A two step procedure to
handle this problem was employed. The first step was to determine 
how to add \carb and empty target data properly 
in order to have the same ratio of heavier nuclei and helium as in the \nh data. 
Using a calculation based on the target thicknesses, densities, and 
window contributions, the background spectrum was calculated as 
$N^{BG}=N^C-(0.331\pm0.008)N^E$, were $N^C$ and $N^E$ are the total 
number of \carb and empty target data, normalized to the same charge. 

\begin{figure*}[here,top]
\centering
\mbox{\subfigure{\epsfig{figure=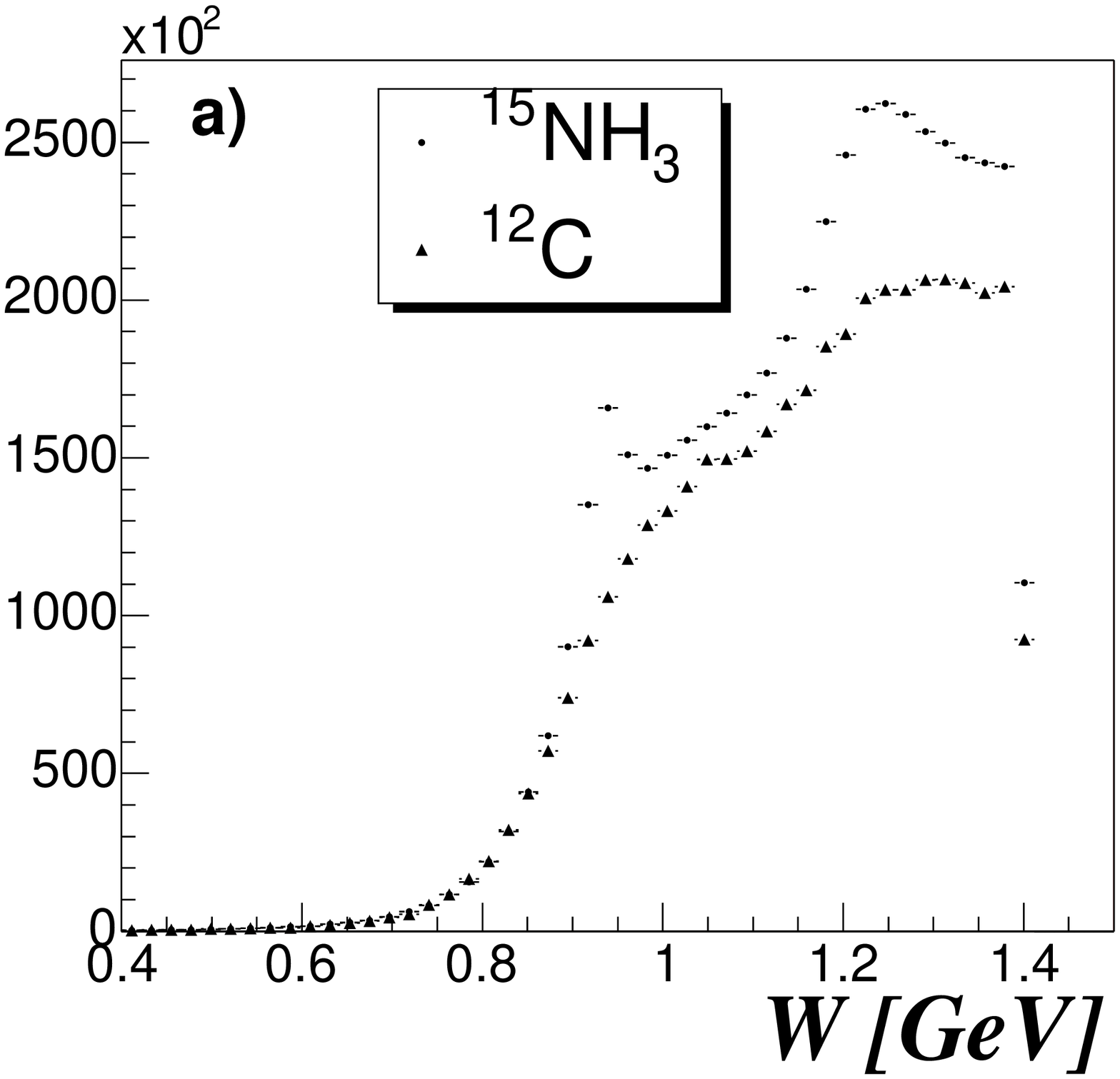,width=2.5in}}\quad
\subfigure{\epsfig{figure=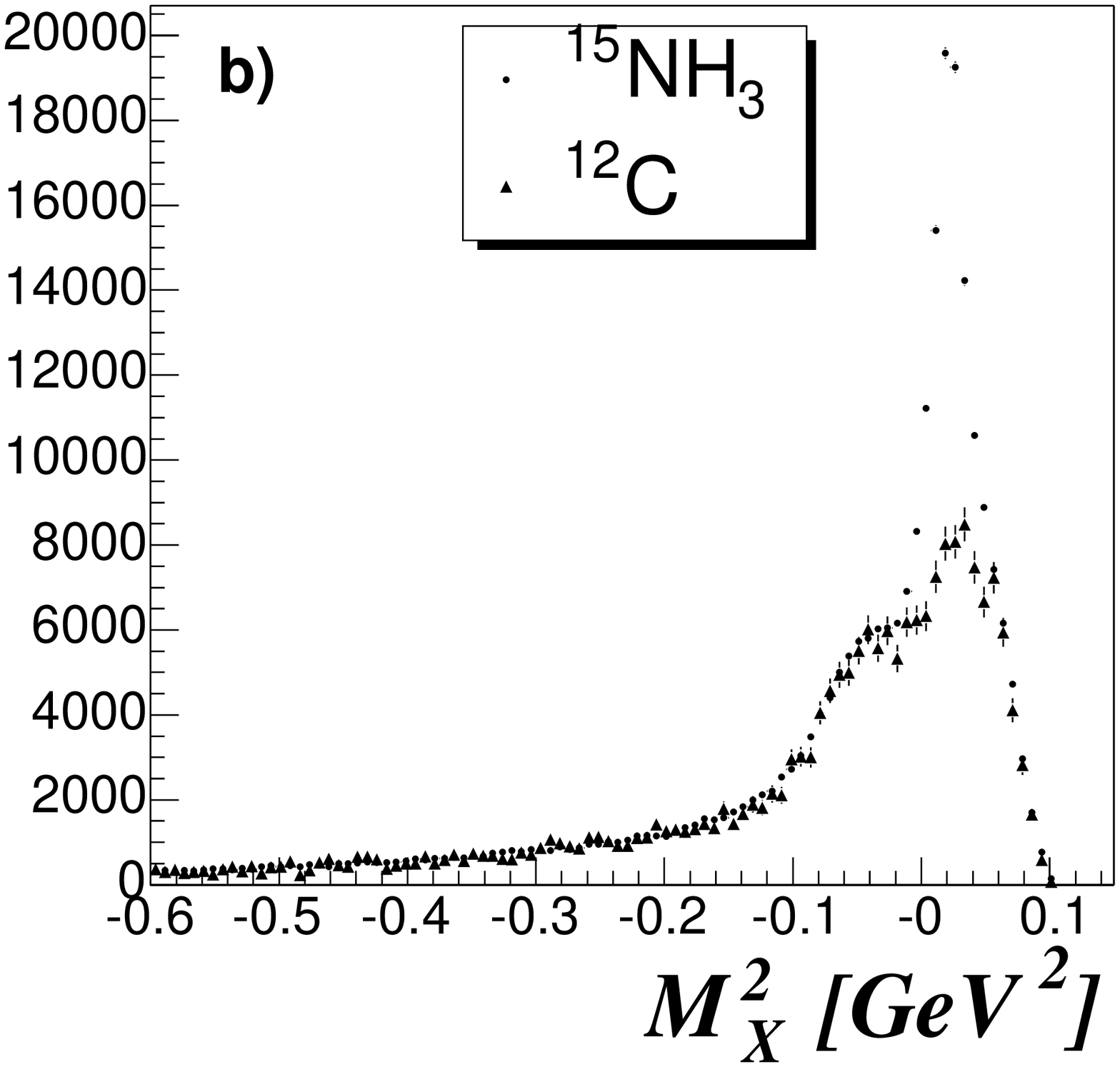,width=2.5in}}}
\caption{{a) Exclusive $W$ spectra for \nh (circles) and \carb (triangles). 
The spectra are normalized to each other using the integrals of the $W$ 
tails in the range $0.6\ \gevm$ to $0.85\ \gevm$. b) Overlay of $M_X^2$ 
spectra for \nh (circles) and \carb (triangles). The \carb was normalized  
using the constant found from the $W$ tail integrals.}}
\label{norm}
\end{figure*}

The second step in the background subtraction was to determine a 
cross-normalization constant $C_{\Delta}$, which allows $N^{BG}$ 
to be equivalent to the rates from $^{15}$N, accounting for the different 
ratio of protons to neutrons between the two backgrounds. 
A constant for the elastic region $C_{el}$ was found as a ratio 
of the integrals of the $W$ tails of the \nh and background data 
from $W=0.6$ to $0.85\ \gevm$ where only events from scattering by 
bound nucleons are present. Figure~\ref{norm}(a) shows the overlay 
of the $W$ spectra of \nh and \carb after normalization by $C_{el}$.  
A correction for higher $W$ was then applied to $C_{el}$ to account 
for rates from scattering off neutrons. 
$C_{\Delta}=\frac{6}{7}\frac{22}{18}C_{el}$ was obtained for the \delt 
region, where $\frac{6}{7}$ is the ratio of protons in \carb and \nitr 
and $\frac{22}{18}$ is based on a Clebsch-Gordan coefficient analysis~\cite{angela}. 
Figure~\ref{norm}(b) shows the overlay of $M_x^2$ for \nh and 
background data after normalization using $C_{\Delta}$. The tails 
where $M_x^2<0$ match as expected since they come only from quasi-elastic scattering 
off the bound nucleons. The technique was later verified using a \nitr target.

\begin{figure}[here,top]
\begin{center}
\includegraphics[width=3.0in]{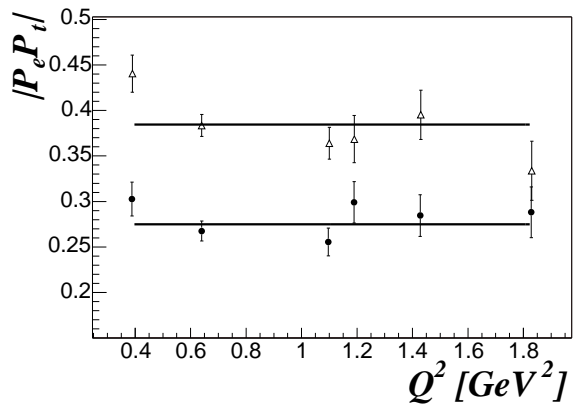}
\end{center}
\caption{The product $|P_eP_t|$ as a function of $Q^2$ for positive (filled circles) and 
negative (open triangles) target polarization runs. The six values for each polarization 
were fitted with a constant in order to obtain the average values 
$P^{}_eP^a_t=0.275 \pm 0.007$ and $P^{}_eP^b_t=-0.385 \pm 0.008$. The values for the $\chi^2$ per degree of freedom 
of the fits were $5.884/5$ and $11.87/5$ respectively (Note suppressed zero).}
\label{pept}
\end{figure}

{\bf Target polarization measurement. }The target polarization was extracted 
by comparing  the well known elastic scattering asymmetry~\cite{elast}
\begin{widetext}
\begin{equation}
A_{theo}=-\frac{\cos\theta_{\gamma}\sqrt{1-\epsilon^2}+(\frac{Q^2}{4M^2})^{-\frac{1}{2}}\sqrt{2\epsilon(1-\epsilon)}\sin\theta_{\gamma}\cos\phi_{\gamma}\frac{G_E}{G_M}}{\epsilon(\frac{Q^2}{4M^2})^{-1}(\frac{G_E}{G_M})^2+1}
\end{equation}
\end{widetext}
with the measured  asymmetry
\begin{equation}
A_{meas}=\frac{\npp-\nmp}{\npp+\nmp}=\frac{P_eP_t\sigma_{et}}{\sigma_0} \equiv P_eP_tA_{theo}.
\end{equation}
The ratio $\frac{G_E}{G_M}$ has been measured in many experiments and it is known within a 3\percent accuracy in the $Q^2$ region of interest~\cite{jones}.
The product of beam and target polarization ($P_eP_t$)  was independently estimated using six $Q^2$ 
bins and then the average value was calculated. Figure~\ref{pept} shows the 
results for the positive ($P^{}_eP^a_t$) and negative target polarization data 
($P^{}_eP^b_t$). These measurements allow one to extract target polarizations $P^a_t$, $P^b_t$ 
by simply taking the ratio of these products and the measured beam 
polarization $P_e$ (see section~\ref{expsetup}).

{\bf Systematic uncertainties. }Several sources of possible systematic 
effects were identified in the analysis procedure. To estimate the size
of these uncertainties, asymmetries were recalculated changing individual parameters 
in the analysis and comparing with the original result. Table~\ref{summsysterr} 
summarizes the systematic uncertainties for $A_{et}$ in the bin 
$0.9\ \gevpsq<Q^2<1.5\ \gevpsq$. Similar values were found for the 
other asymmetries and $Q^2$ bins. The overall systematic uncertainty is 
on the order of 5$\%$, which is much smaller than the statistical 
uncertainty for the measured asymmetries.
\begin{table}[here,top]
\caption{\label{summsysterr} Summary of the systematic uncertainties for the 
asymmetry $A_{et}$ %as a function of $\phi^*$ 
for $0.9\ \gevpsq<Q^2<1.5\ \gevpsq$.}
\begin{ruledtabular}
\begin{tabular}{ c c } 
{\sl Systematic uncertainty source}     & Systematic uncertainty (\percent)  \\ \hline
Carbon normalization            &4.2            \\ 
$P_eP_t$                &2.3                    \\ 
$P_e$                   &1.3                    \\ 
\hel background contribution            &3.3                    \\ 
%Norm const $d$         &$<$0.1                 \\ 
\end{tabular}
\end{ruledtabular}
\end{table}

{\bf Radiative corrections. }
Radiative corrections were estimated using a generalization of the 
Mo-Tsai formulation~\cite{Mo-Tsai}. In particular the corrections 
were obtained by comparing Monte Carlo generated radiative and non-radiative 
events.
The regions with zero acceptance existing in the data were incorporated 
in the Monte Carlo in order to improve the model representation 
of the data. The difference between asymmetries calculated with radiative 
and non-radiative events revealed that radiative corrections 
influence the data by at most a few percent.

\section{Results} 
\begin{figure}[here,top]
\begin{center}
\includegraphics[width=2.5in]{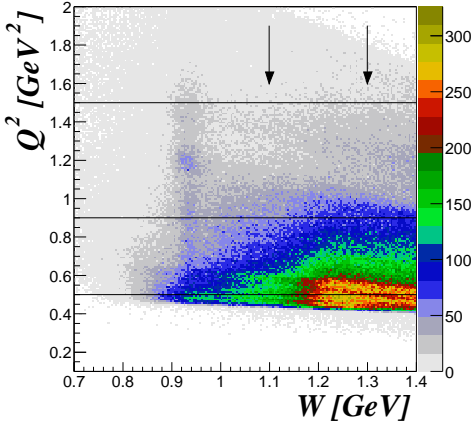}
\end{center}
\caption{(color) $Q^2\ vs\ W$. In the \delt region the accessible range in $Q^2$ 
is from $0.4\ \gevpsq$ to $1.5\ \gevpsq$. The horizontal lines delineate 
the two intervals of $Q^2$ in which the data were divided.}
\label{kin}
\end{figure}

\begin{figure*}[ht]
\begin{center}
\epsfig{file=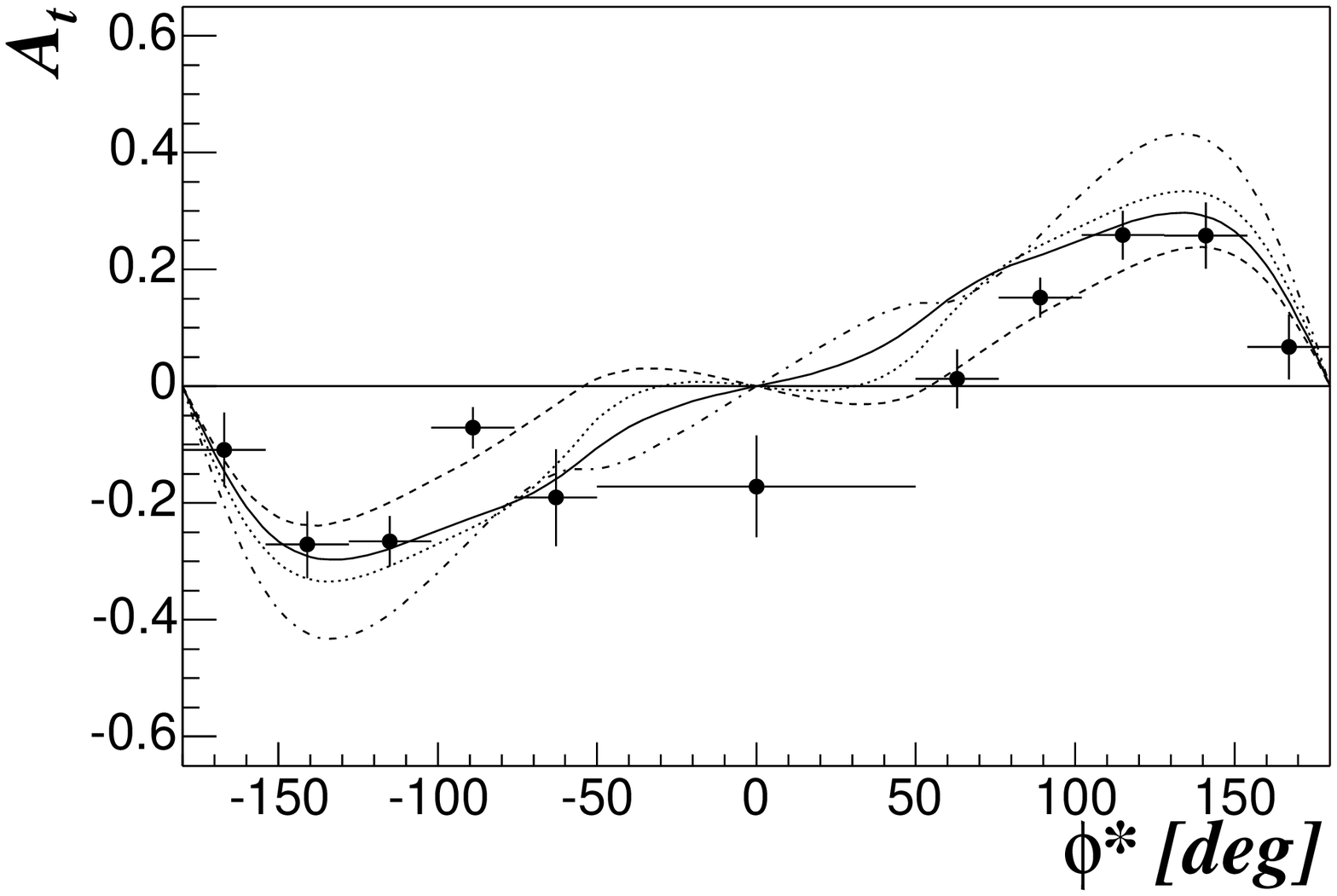,width=17pc}
\epsfig{file=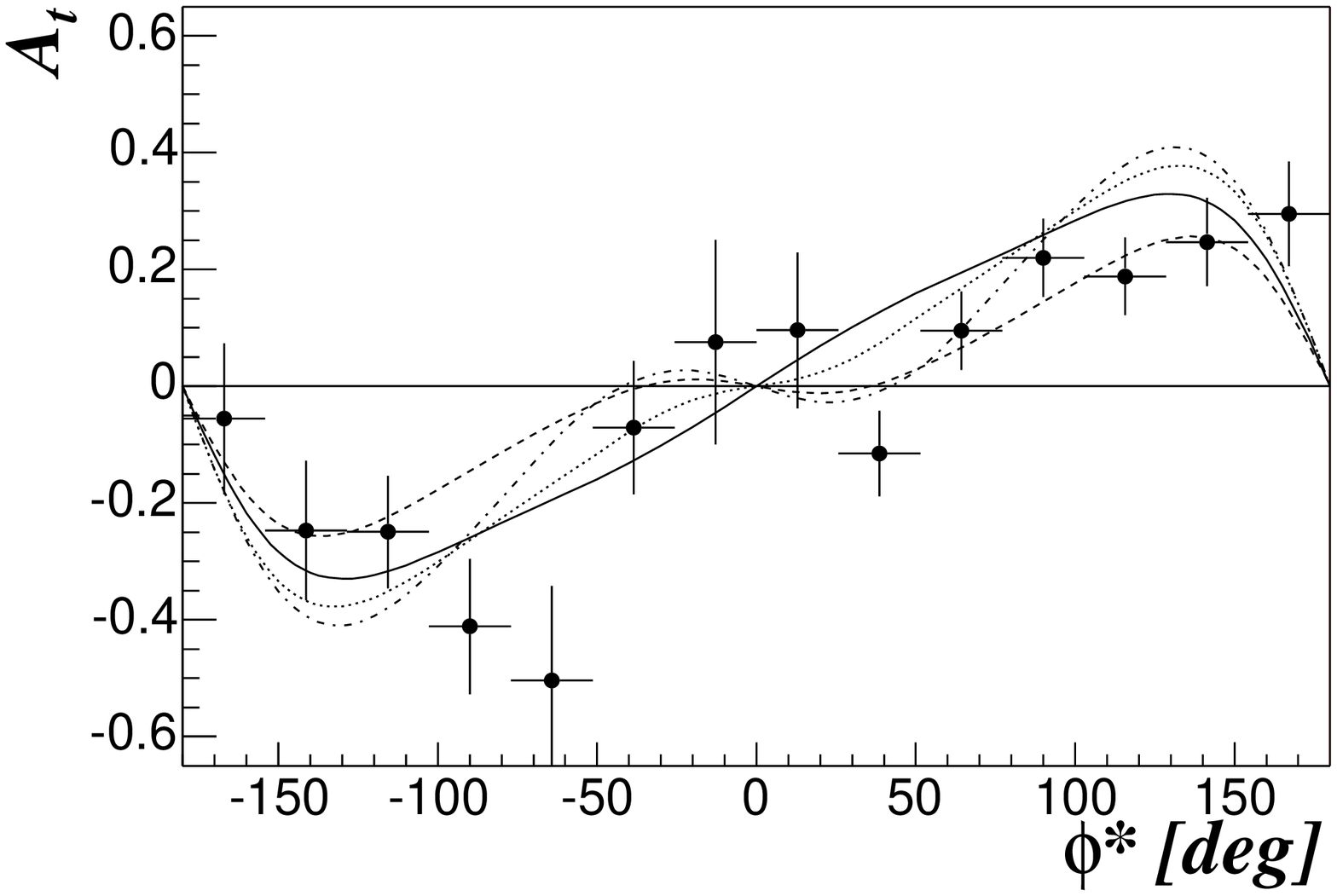,width=17pc}

\epsfig{file=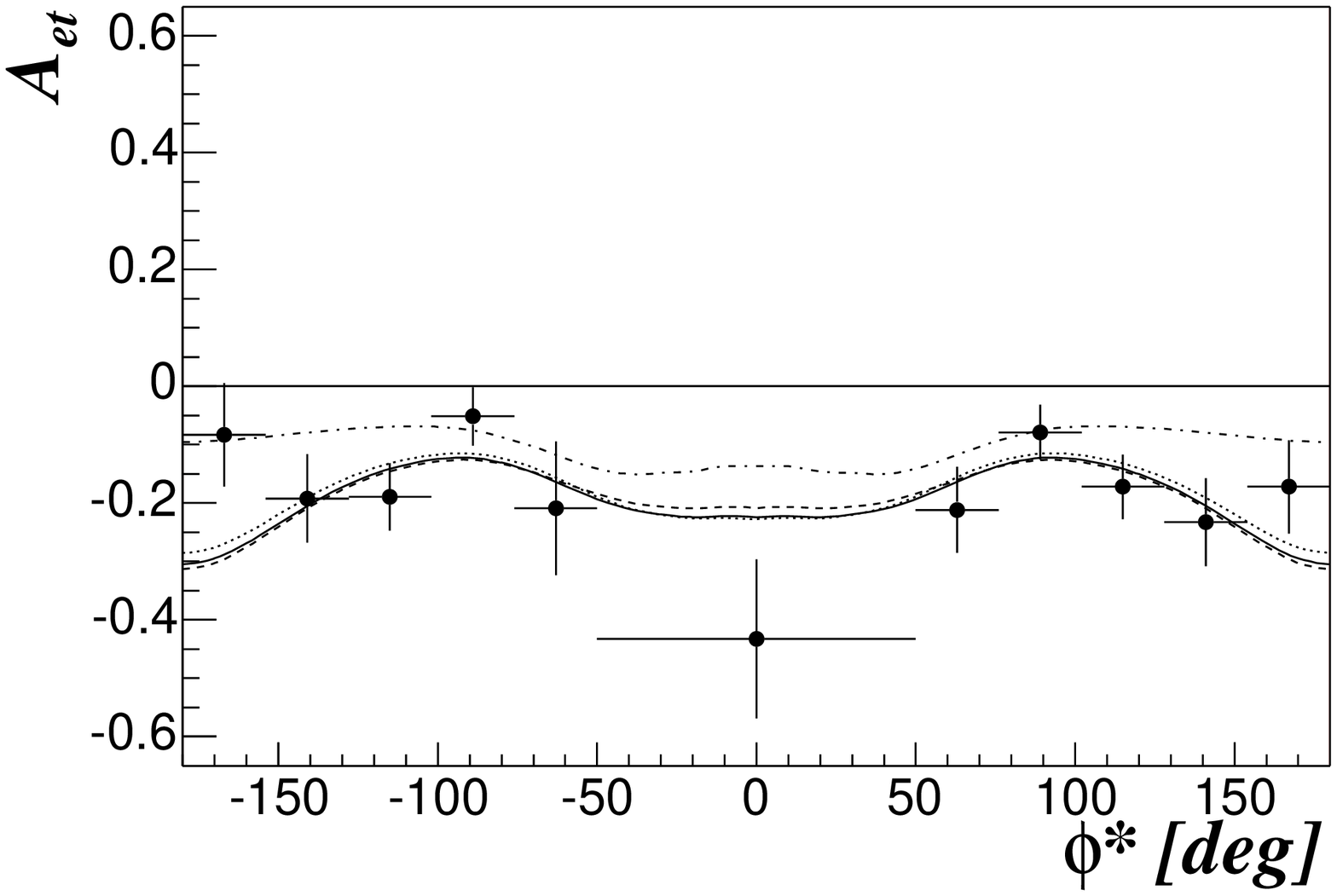,width=17pc}
\epsfig{file=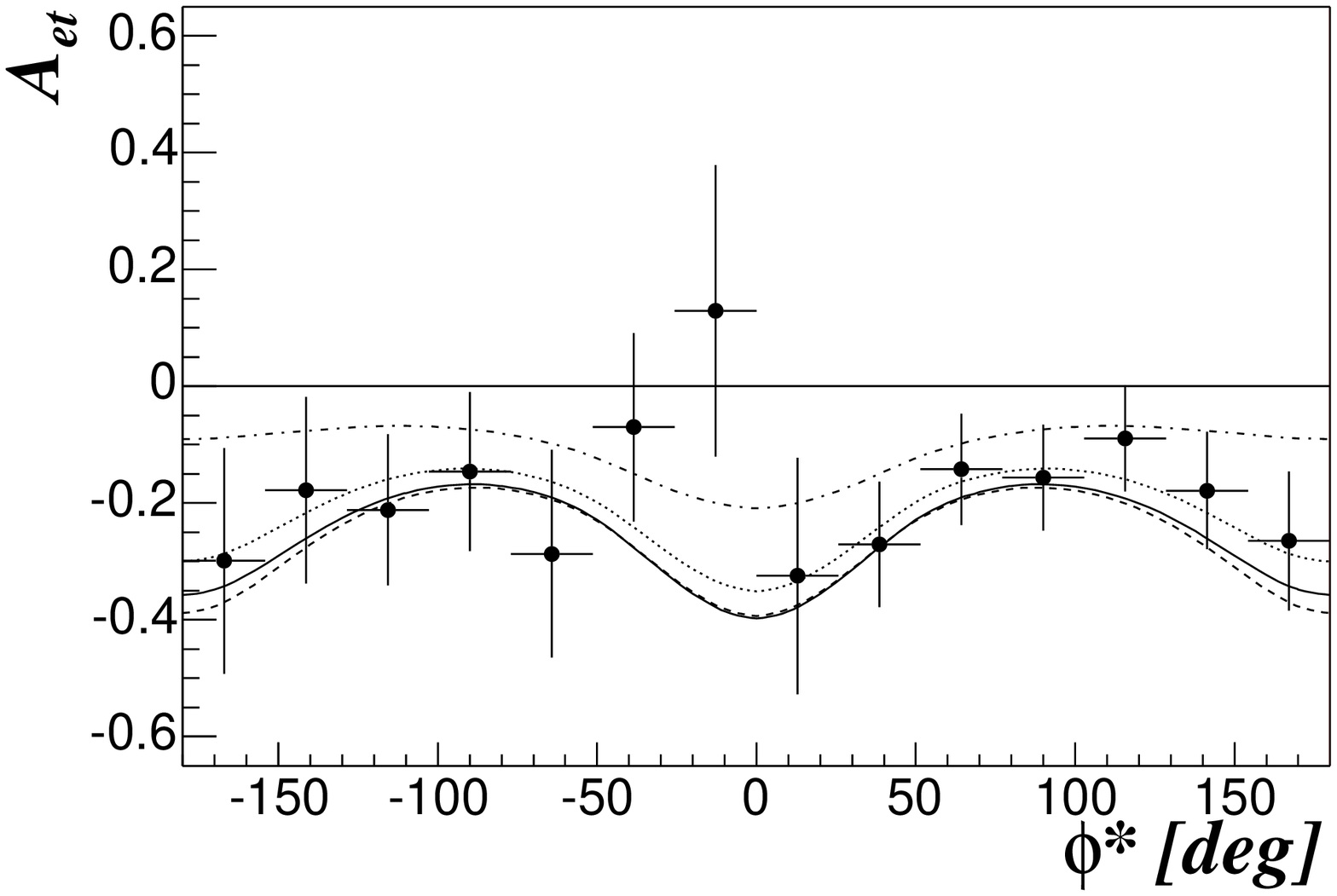,width=17pc}

\caption{Asymmetries $A_{t}$ and $A_{et}$ as a function of the center-of-mass
angle of the pion $\phi^{*}$  integrated over $\cos\theta^{*}$ for 
$0.5\ \gevpsq<Q^2<0.9\ \gevpsq$ (left) and 
$0.9\ \gevpsq<Q^2<1.5\ \gevpsq$ (right). The curves represent 
the predictions from the MAID2000 model (solid), 
Davidson-Mukhopadhyay model 
(dash-dotted), Sato-Lee model (dashed), and DMT model (dotted). 
}
\label{asymph}
\end{center}
\end{figure*}

\begin{figure*}[ht]
\begin{center}
\epsfig{file=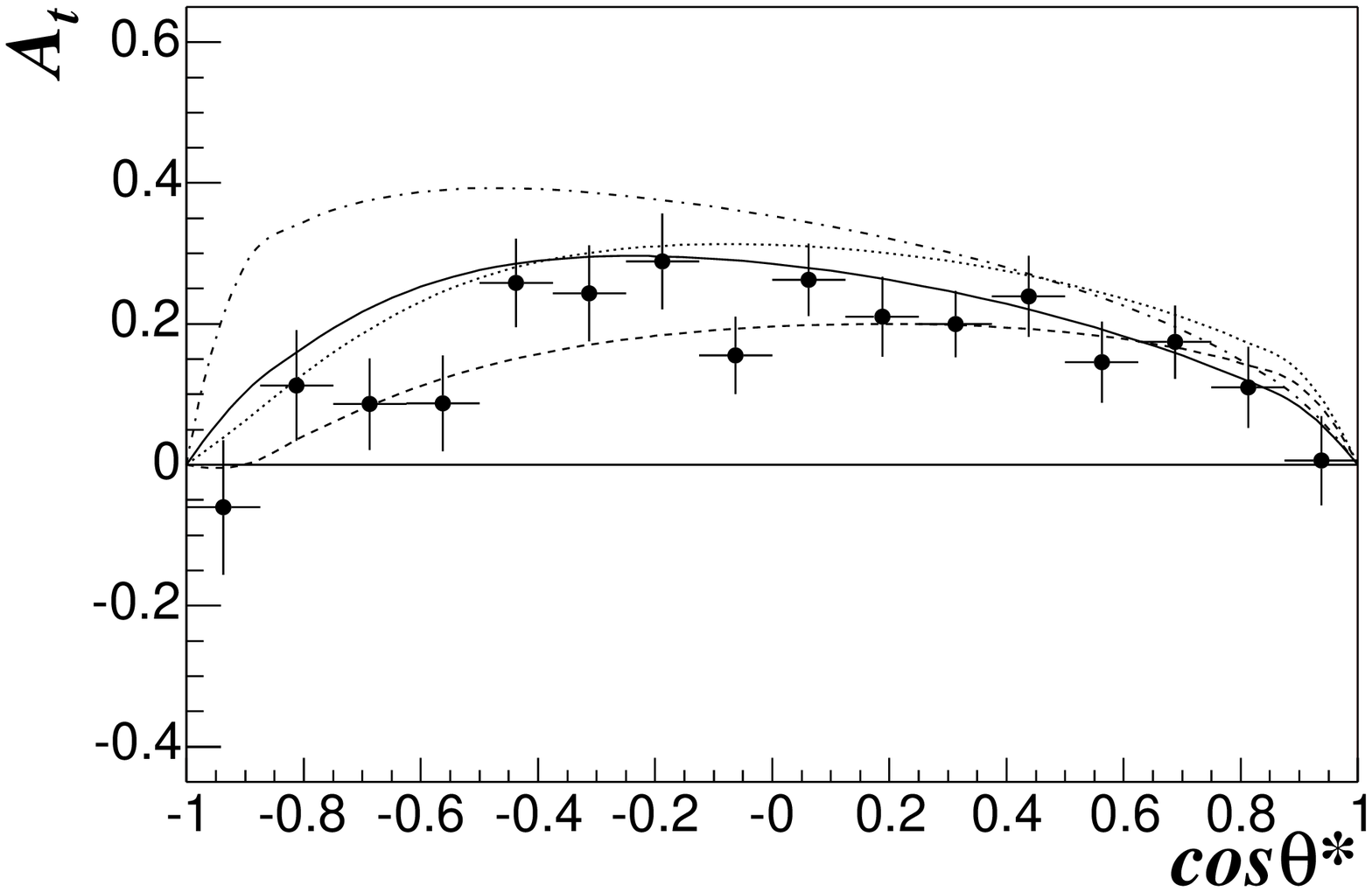,width=17pc}
\epsfig{file=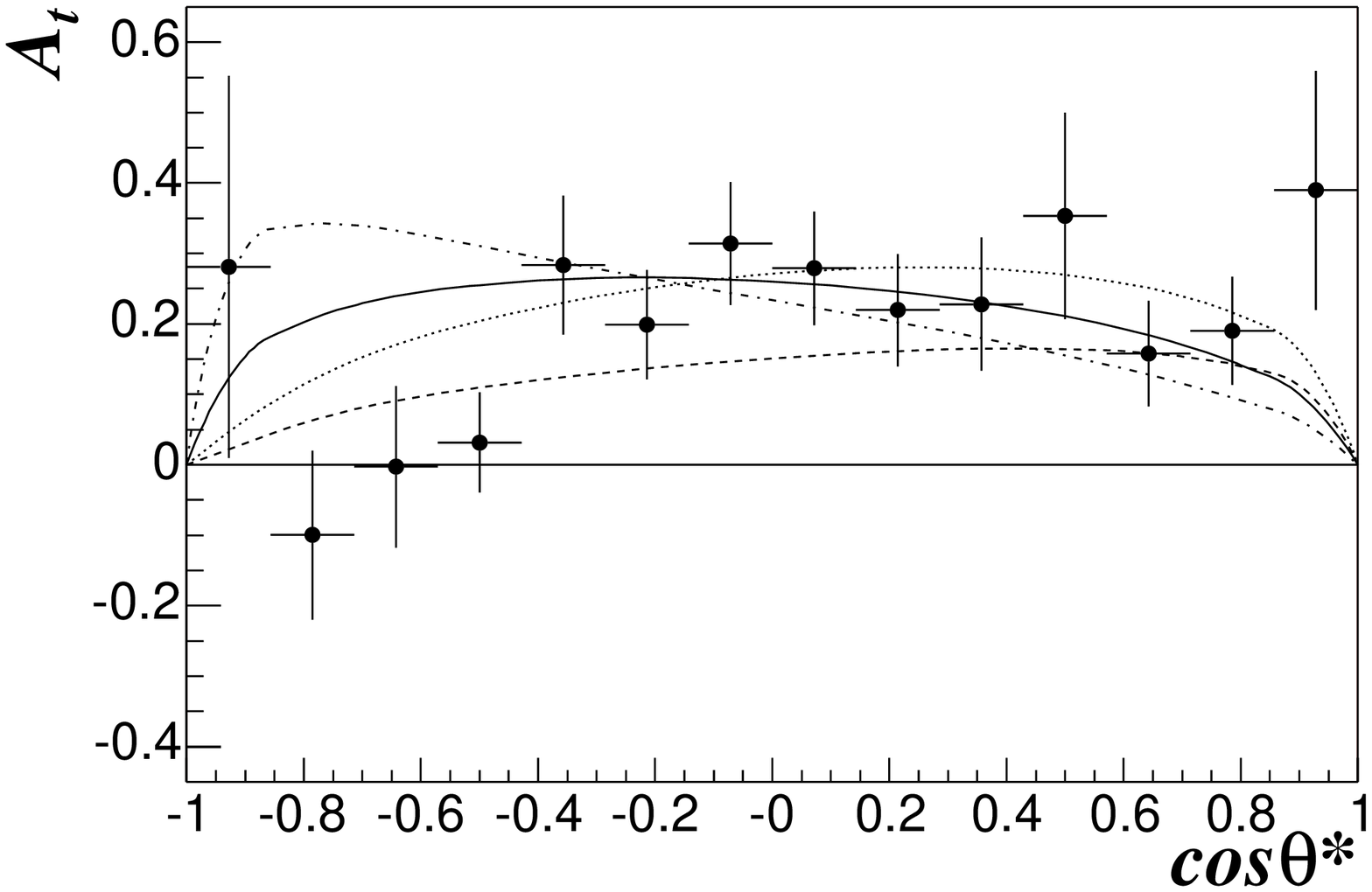,width=17pc}

\epsfig{file=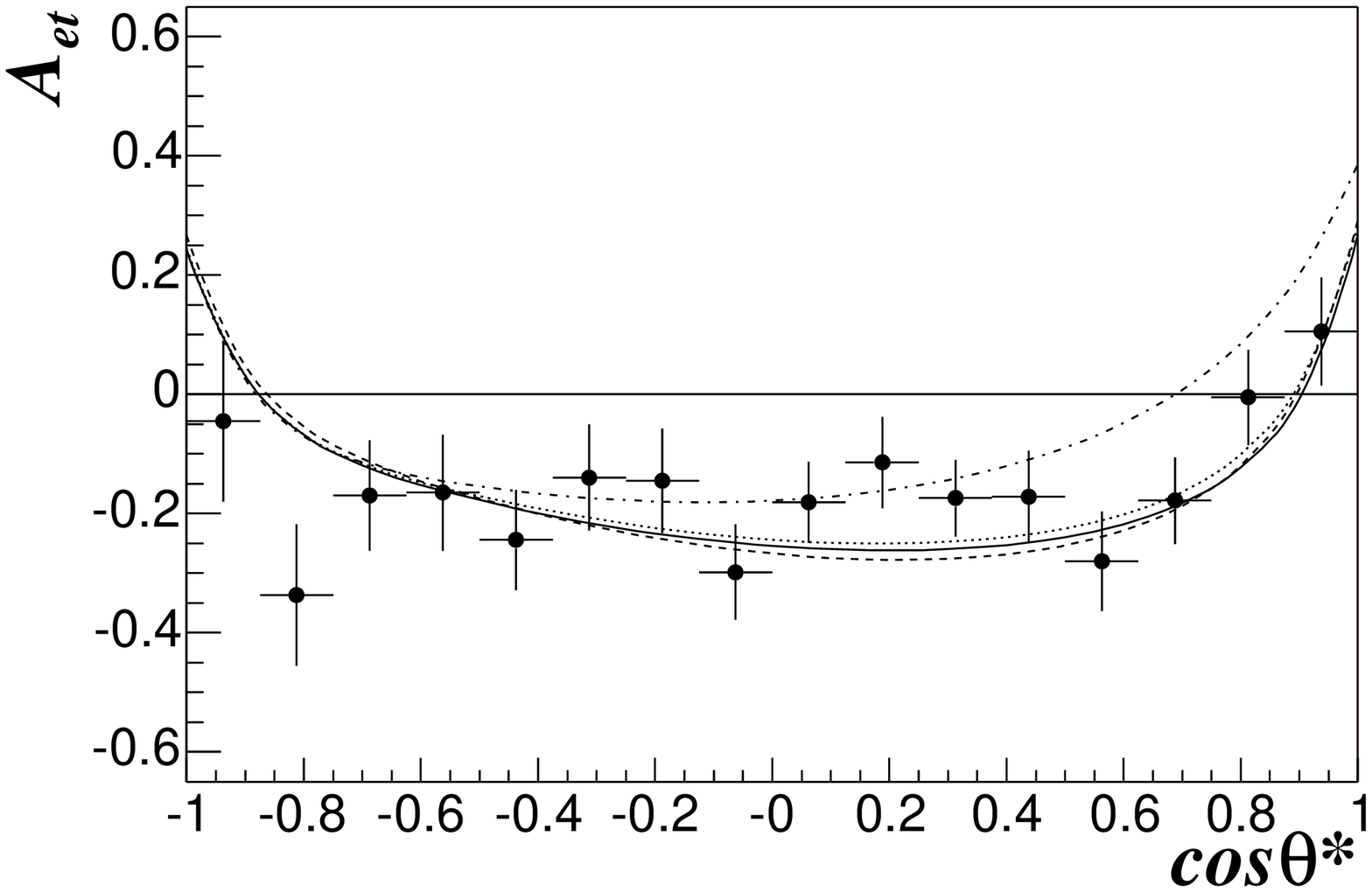,width=17pc}
\epsfig{file=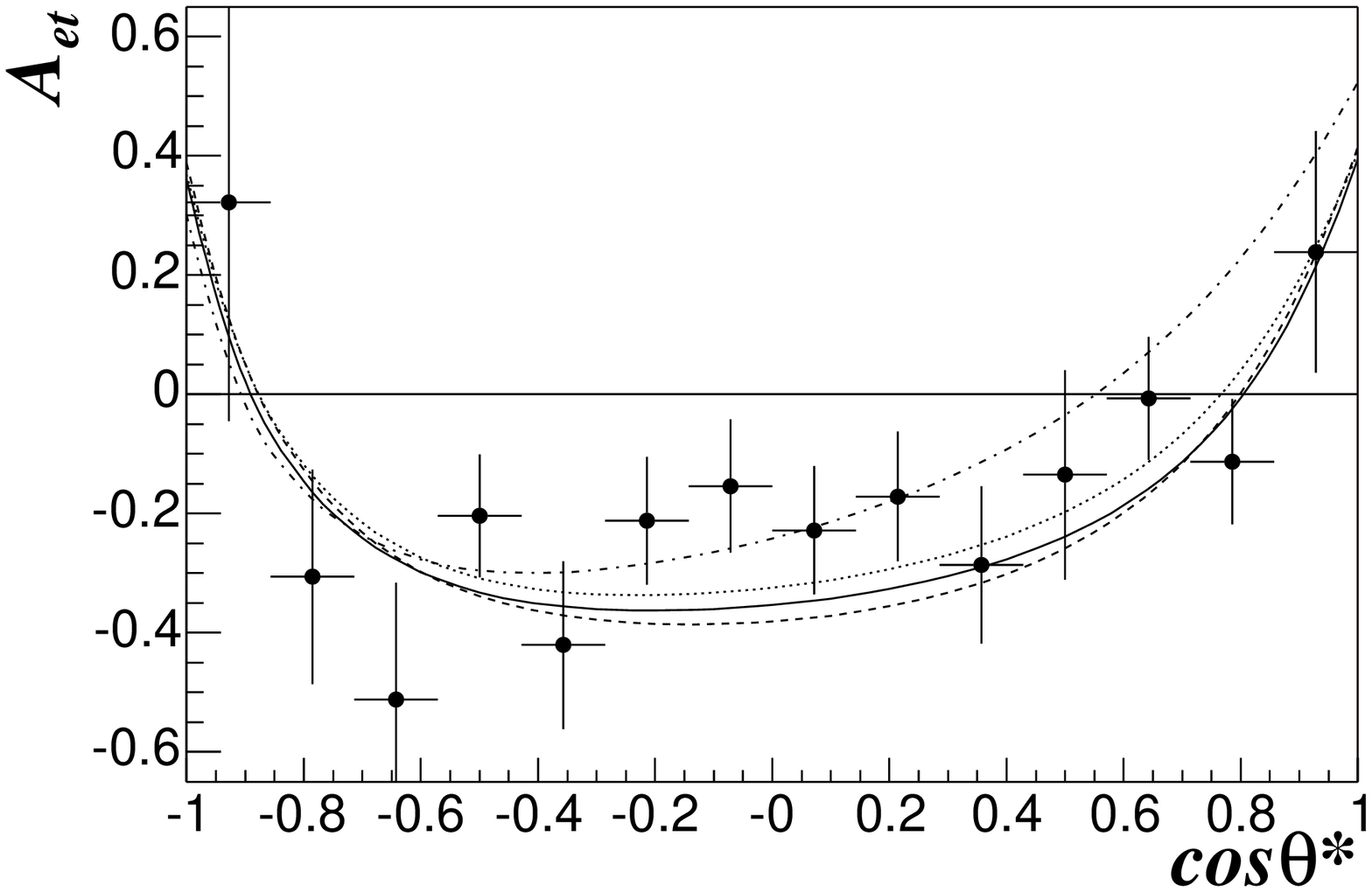,width=17pc}
\caption{Asymmetries $A_{t}$ and $A_{et}$ as a function of the 
center-of-mass angle of the pion $\cos \theta^{*}$  integrated  
over $0\deg <\phi^{*}<180\deg$
and $-180\deg <\phi^{*}<180\deg$, respectively
for $0.5\ \gevpsq<Q^2<0.9\ \gevpsq$ (left) and 
$0.9\ \gevpsq<Q^2<1.5\ \gevpsq$ (right). The curves represent 
the predictions from the MAID2000 model (solid), 
Davidson-Mukhopadhyay model 
(dash-dotted), Sato-Lee model (dashed), and DMT model (dotted). Note that the complete data set contributes to the determination
of $A_t$ by making use of the symmetry of $\sigma_t$ with respect to $\phi^*$. This was achieved by integrating 
the terms for $\sigma_t$ in equations~(\ref{forasymdef}) 
 for positive and negative $\phi^*$ separately and then adding the two results with opposite sign. 
Also note that the results for the lower $Q^2$ bin are affected by the zero acceptance region (see Fig~\ref{acc}).}
\label{asymth}
\end{center}
\end{figure*}

\begin{figure*}[ht]
\begin{center}
\epsfig{file=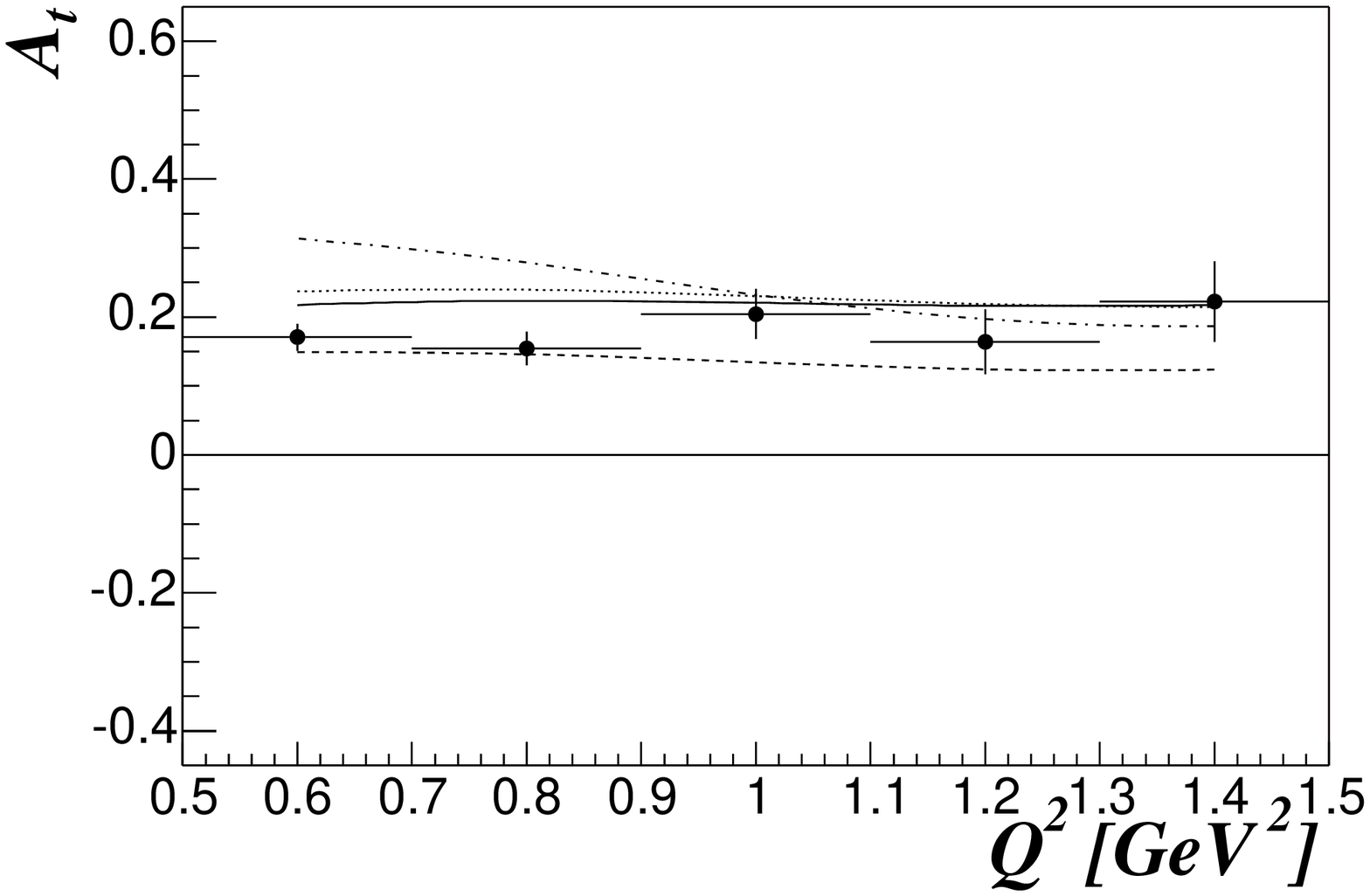,width=17pc}
\epsfig{file=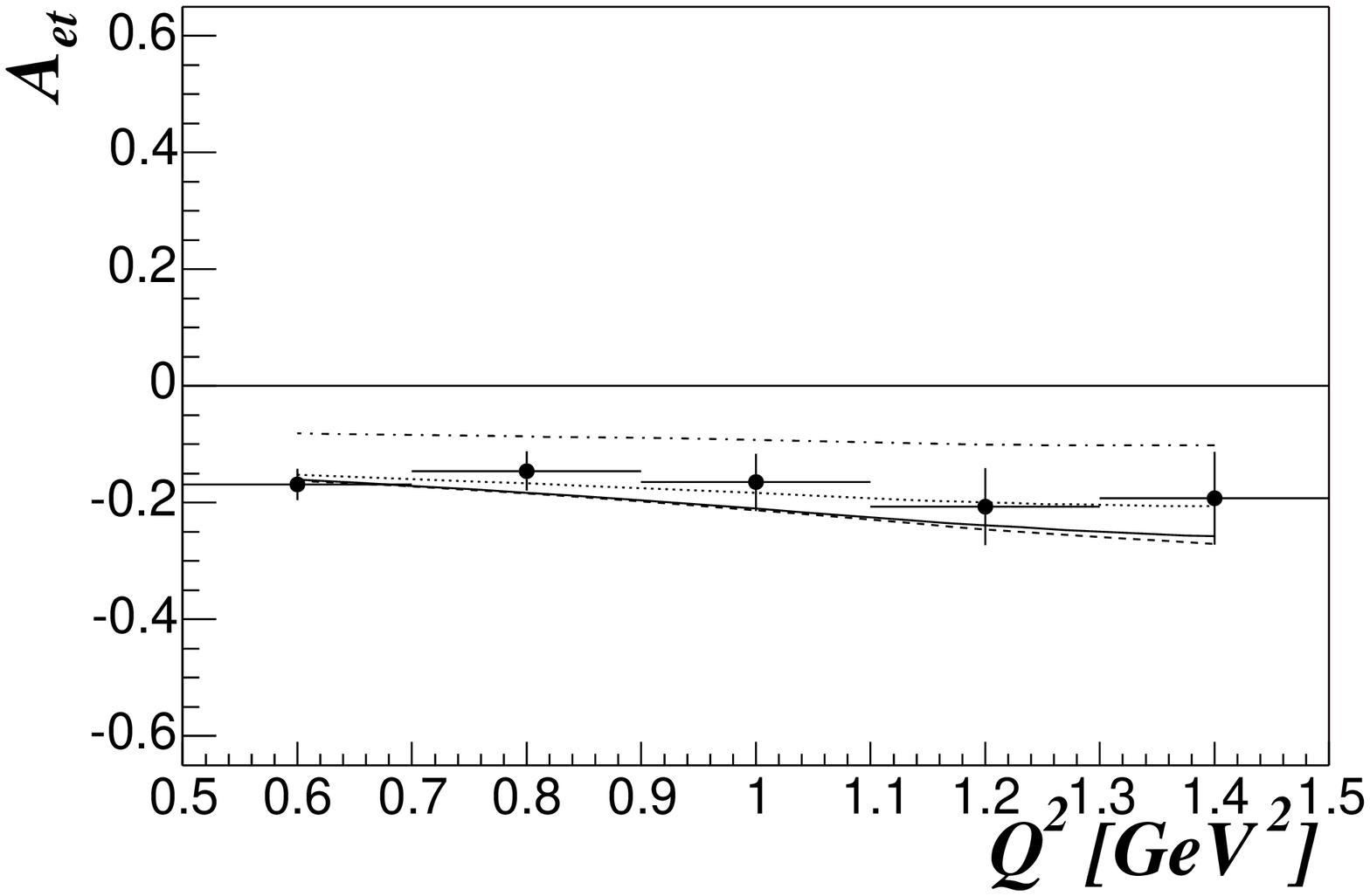,width=17pc}

\caption{Asymmetries $A_{t}$ and $A_{et}$ as a function of the 
momentum transfer, $Q^2$, integrated over $\cos\theta^{*}$ and $0\deg <\phi^{*}<180\deg$
and $-180\deg <\phi^{*}<180\deg$, respectively. The curves represent the predictions 
from the MAID2000 model (solid black), 
Davidson-Mukhopadhyay model (dash-dotted), Sato-Lee model (dashed), and DMT model (dotted).}
\label{asymQ}
\end{center}
\end{figure*}

Data for a beam energy of $2.565\ \gev$, within the \delt region 
($1.1\ \gevm<W<1.3\ \gevm$), span a range in momentum transfer, $Q^{2}$, from 
$0.4\ \gevpsq$ to $1.5\ \gevpsq$, as can be seen in Figure~\ref{kin}.  
The data were divided in two $Q^2$ bins, $0.5\ \gevpsq<Q^2<0.9\ \gevpsq$ 
and $0.9\ \gevpsq<Q^2<1.5\ \gevpsq$, and the asymmetries $A_{t}$  and $A_{et}$ 
were extracted according to the definitions in equations~(\ref{forasymdef}) as a function 
of the 
angle of the pion in the center-of-mass $\phi^*$, integrated over $\cos\theta^*$, and conversely
as a function of $\cos\theta^*$, integrated over $\phi^*$. 
The $Q^2$ dependences integrated over $\phi^*$ and 
$\cos\theta^*$ were extracted as well. 
The results are shown in Figures~\ref{asymph},~\ref{asymth}, 
and~\ref{asymQ} and listed in 
Table~\ref{tableresults} -~\ref{tableresults5}.
The beam asymmetry was not extracted because it could not
be separated from the background stemming from \delt $\rightarrow \pi^-p$ that is 
produced by the scattering off neutrons in the $^{15}$N.

According to equation~(\ref{forsigma}) the asymmetries depend on 
$\sin \phi^{*}$, $\cos \phi^{*}$, $\sin 2\phi^{*}$ and $\cos 2\phi^{*}$ giving 
a well defined functional dependence in $\phi^{*}$ that is model independent, and  
the data were found to agree with this expectation.
The target 
asymmetry was found to be an odd function and a fit to the function 
$\frac{A\cos \phi^{*} \sin \phi^{*} +B\sin \phi^{*}+C\sin^3 \phi^{*}}{D+E\cos\phi^{*}+F\cos 2\phi^{*}}$ 
gave $\chi^2$ per degree of freedom ($ndf$) values of 7.9/9 and 15.4/9 for the low and high $Q^2$ 
bin respectively. The double spin asymmetry was fitted with the even function 
$\frac{A +B\cos \phi^{*}+C\cos^2 \phi^{*}}{D+E\cos\phi^{*}+F\cos 2\phi^{*}}$ and the values 
$\chi^2/ndf=4.4/9$ for $0.5\ \gevpsq<Q^2<0.9\ \gevpsq$ and 4.8/7 for
$0.9\ \gevpsq<Q^2<1.5\ \gevpsq$ were found. 

\begin{table}[here,top]
\caption{\label{tableresults}Asymmetries $A_{t}$ and $A_{et}$ 
as a function of the center-of-mass angle of the pion $\phi^{*}$ integrated over $\cos\theta^{*}$ at low $Q^2$. The uncertainties listed are statistical and systematic, respectively.}
\begin{ruledtabular}
\begin{tabular}{c r r}
\multicolumn{3}{c}{$0.5 \ \gevpsq <Q^{2}<0.9 \ \gevpsq$}\\ \hline
$\phi^{*} \ [\deg]$ & $A_t$ & $A_{et}$ \\ \hline
-167.     & $-0.108\pm0.063\pm0.008$ & $-0.083\pm0.088\pm0.006$\\ 
-141.     &  $-0.271\pm0.058\pm0.018$ & $-0.192\pm0.076\pm0.014$\\ 
-115.     & $-0.266\pm0.044\pm0.016$ & $-0.189\pm0.058\pm0.012$\\ 
-89.      & $-0.071\pm0.036\pm0.006$ & $-0.052\pm0.050\pm0.003$\\ 
-63.      & $-0.191\pm0.083\pm0.012$ & $-0.209\pm0.115\pm0.017$\\ 
0.        & $-0.171\pm0.087\pm0.028$ & $-0.433\pm0.136\pm0.026$\\ 
63.       & $0.013\pm0.051\pm0.004$ & $-0.211\pm0.074\pm0.016$\\ 
89.       & $0.152\pm0.034\pm0.011$ & $-0.079\pm0.047\pm0.004$\\ 
115.      & $0.259\pm0.042\pm0.017$ & $-0.172\pm0.055\pm0.012$\\ 
141.      & $0.258\pm0.056\pm0.021$ & $-0.232\pm0.075\pm0.020$\\ 
167.      & $0.067\pm0.056\pm0.006$  & $-0.172\pm0.080\pm0.014$\\ 
\end{tabular}
\end{ruledtabular}
\end{table}

\begin{table}[here,top]
\caption{\label{tableresults2}Asymmetries $A_{t}$ and $A_{et}$
as a function of the center-of-mass angle of the pion $\cos\theta^{*}$ integrated over $\phi^{*}$ at low $Q^2$. The uncertainties listed are statistical and systematic, respectively. Please note that the results in this table are affected by the zero acceptance region (see Fig~\ref{acc}).}
\begin{ruledtabular}
\begin{tabular}{c r r } 
\multicolumn{3}{ c }{$0.5 \ \gevpsq <Q^{2}<0.9 \ \gevpsq$}\\ \hline
 &  $0 \deg<\phi^*<180 \deg$&  $-180 \deg<\phi^*<180 \deg$\\ \hline
$\cos\theta^{*}$ & $A_t$ & $A_{et}$ \\ \hline
$-0.938$      & $-0.061\pm0.096\pm0.038 $ & $-0.045\pm0.135\pm0.008$\\
$-0.812$      & $ 0.113\pm0.078\pm0.008$ & $-0.336\pm0.119\pm0.044$\\
$-0.688$      & $ 0.086\pm0.065\pm0.003$ & $-0.170\pm0.093\pm0.015$\\
$-0.562$      & $ 0.087\pm0.068\pm0.002$ & $-0.165\pm0.097\pm0.014$\\
$-0.438$      & $ 0.258\pm0.063\pm0.018$ & $-0.244\pm0.084\pm0.021$\\
$-0.312$      & $ 0.243\pm0.068\pm0.032$ & $-0.139\pm0.089\pm0.012$\\
$-0.188$      & $ 0.289\pm0.068\pm0.017$ & $-0.145\pm0.088\pm0.011$\\
$-0.062$      & $ 0.155\pm0.055\pm0.015$ & $-0.298\pm0.080\pm0.017$\\
$0.062$       & $ 0.262\pm0.051\pm0.013$ & $-0.181\pm0.068\pm0.010$\\
$0.188$       & $ 0.210\pm0.057\pm0.029$ & $-0.114\pm0.077\pm0.006$\\
$0.312$       & $ 0.200\pm0.047\pm0.011$ & $-0.174\pm0.064\pm0.010$\\
$0.438$       & $ 0.239\pm0.058\pm0.008$ & $-0.172\pm0.077\pm0.010$\\
$0.562$       & $ 0.146\pm0.058\pm0.014$ & $-0.280\pm0.084\pm0.018$\\
$0.688$       & $ 0.174\pm0.052\pm0.021$ & $-0.178\pm0.072\pm0.012$\\
$0.812$       & $ 0.110\pm0.058\pm0.004$ & $-0.005\pm0.080\pm0.002$\\
$0.938$       & $ 0.006\pm0.063\pm0.005$ & $ 0.106\pm0.091\pm0.010$\\ 
\end{tabular}
\end{ruledtabular}
\end{table}

\begin{table}[here,top]
\caption{\label{tableresults3}Asymmetries $A_{t}$ and $A_{et}$ as
a function of the center-of-mass angle of the pion $\phi^{*}$ integrated over $\cos\theta^{*}$ at high $Q^2$. The uncertainties listed are statistical and systematic, respectively. }
\begin{ruledtabular}
\begin{tabular}{c r r } 
\multicolumn{3}{ c }{$0.9 \ \gevpsq <Q^{2}<1.5 \ \gevpsq $}\\ \hline
$\phi^{*} \ [\deg]$ & $A_t$ & $A_{et}$ \\ \hline
$-167.1$      & $-0.056\pm0.129\pm0.004$ & $-0.299\pm0.194\pm0.020$\\
$-141.4$      & $-0.247\pm0.119\pm0.015$ & $-0.178\pm0.160\pm0.012$\\
$-115.7$      & $-0.250\pm0.096\pm0.015$ & $-0.212\pm0.130\pm0.013$\\
$-90.0$       & $-0.411\pm0.116\pm0.026$ & $-0.146\pm0.136\pm0.011$\\
$-64.3$       & $-0.504\pm0.162\pm0.053$ & $-0.287\pm0.178\pm0.031$\\
$-38.6$       & $-0.071\pm0.115\pm0.005$ & $-0.070\pm0.162\pm0.006$\\
$-12.9$       & $0.076\pm0.176\pm0.011$  & $0.129\pm0.249\pm0.021$\\
$12.9$        & $0.096\pm0.133\pm0.009$  & $-0.325\pm0.202\pm0.033$\\
$38.6$        & $-0.115\pm0.074\pm0.009$ & $-0.271\pm0.107\pm0.020$\\
$64.3$        & $0.095\pm0.067\pm0.006$  & $-0.142\pm0.095\pm0.009$\\
$90.0$        & $0.220\pm0.067\pm0.012$  & $-0.156\pm0.090\pm0.009$\\
$115.7$       & $0.189\pm0.067\pm0.011$  & $-0.089\pm0.091\pm0.005$\\
$141.4$       & $0.247\pm0.076\pm0.020$  & $-0.179\pm0.101\pm0.014$\\
$167.1$       & $0.295\pm0.089\pm0.015$  & $-0.265\pm0.119\pm0.014$\\
\end{tabular}
\end{ruledtabular}
\end{table}

\begin{table}[here,top]
\caption{\label{tableresults4}Asymmetries $A_{t}$ and $A_{et}$
as a function of the center-of-mass angle of the pion $\cos\theta^{*}$ integrated over $\phi^{*}$ at high $Q^2$. The uncertainties listed are statistical and systematic, respectively. }
\begin{ruledtabular}
\begin{tabular}{c r r } 
\multicolumn{3}{ c }{$0.9 \ \gevpsq <Q^{2}<1.5 \ \gevpsq$}\\ \hline
 &  $0 \deg<\phi^*<180 \deg$&  $-180 \deg<\phi^*<180 \deg$\\ \hline
$\cos\theta^{*}$ & $A_t$ & $A_{et}$ \\ \hline
$-0.929$        & $0.281\pm0.271\pm0.002$  & $0.322\pm0.366\pm0.096$\\
$-0.786$        & $-0.100\pm0.120\pm0.008$ & $-0.306\pm0.180\pm0.031$\\
$-0.643$        & $-0.003\pm0.115\pm0.002$ & $-0.512\pm0.196\pm0.049$\\
$-0.500$        & $0.032\pm0.071\pm0.008$  & $-0.203\pm0.103\pm0.016$\\
$-0.357$        & $0.283\pm0.099\pm0.012$  & $-0.420\pm0.141\pm0.029$\\
$-0.214$        & $0.199\pm0.078\pm0.009$  & $-0.212\pm0.107\pm0.011$\\
$-0.071$        & $0.314\pm0.087\pm0.015$  & $-0.154\pm0.112\pm0.007$\\
$0.071$         & $0.279\pm0.081\pm0.009$  & $-0.228\pm0.108\pm0.014$\\
$0.214$         & $0.220\pm0.080\pm0.014$  & $-0.171\pm0.109\pm0.008$\\
$0.357$         & $0.228\pm0.094\pm0.019$  & $-0.286\pm0.132\pm0.016$\\
$0.500$         & $0.354\pm0.147\pm0.014$  & $-0.135\pm0.176\pm0.011$\\
$0.643$         & $0.158\pm0.075\pm0.008$  & $-0.007\pm0.103\pm0.002$\\
$0.786$         & $0.190\pm0.077\pm0.008$  & $-0.113\pm0.105\pm0.007$\\
$0.929$         & $0.390\pm0.169\pm0.008$  & $0.239\pm0.203\pm0.025$\\
\end{tabular}
\end{ruledtabular}
\end{table}

\begin{table}[here,top]
\caption{\label{tableresults5}Asymmetries $A_{t}$ and $A_{et}$
as a function of the momentum transfer, $Q^2$, integrated over $\phi^{*}$ and $\cos\theta^{*}$. The uncertainties listed are statistical and systematic, respectively. }
\begin{ruledtabular}
\begin{tabular}{c r r } 
\multicolumn{3}{c}{$-1<\cos\theta^*<1$ }\\ \hline
 &  $0 \deg<\phi^*<180 \deg$&  $-180 \deg<\phi^*<180 \deg$\\ \hline
$Q^2 \ [\gevpsq]$ & $A_t$ & $A_{et}$ \\ \hline
0.600     & $0.171\pm0.019\pm0.014$ & $-0.169\pm0.027\pm0.012$\\ 
0.800     & $0.154\pm0.024\pm0.007$ & $-0.146\pm0.033\pm0.010$\\ 
1.000     & $0.205\pm0.036\pm0.008$ & $-0.165\pm0.049\pm0.011$\\ 
1.200     & $0.164\pm0.047\pm0.011$ & $-0.207\pm0.066\pm0.012$\\ 
1.400     & $0.223\pm0.059\pm0.017$ & $-0.192\pm0.079\pm0.013$\\ 
\end{tabular}
\end{ruledtabular}
\end{table}

{\bf Comparison with models.} As noted in the introduction, comparisons
of the present results with four theoretical approaches were carried out.
These include MAID2000~\cite{maidref} (MAID), 
an effective Lagrangian model~\cite{davidson} (DM) and the dynamical models 
of SL~\cite{lee,lee2} and DMT~\cite{dmt}. 

{\bf {\boldmath $\chi^2$} comparison. } All the models predict  
the correct sign and the correct order of 
magnitude, but do not yield equally good overall fits to the data.
A simultaneous $\chi^{2}$ 
comparison of all angular distributions, as well as the $Q^{2}$ distributions 
was performed to establish quantitatively 
which model gives the best description of the data. A $\chi^{2}$ comparison
for subsets of the experimental 
distributions was performed as well to understand the model sensitivity 
to the different asymmetries. In order for a $\chi^2$ comparison to be 
made, the model prediction was disregarded where the  
acceptance was zero. 

The $\chi^2$ was defined as:
\begin{equation}
\chi^{2} = \sum_{i} \frac{(x_{i}^{\rm data} - x_{i}^{\rm model})^{2}}{(\sigma_{i}^{\rm data})^{2}},
\end{equation}
where $x_{i}^{\rm data}$ is the value of each experimental point 
for all the asymmetries and $x_{i}^{\rm model}$ is the corresponding 
value of the theoretical prediction. Since the model is given without 
errors, only the experimental uncertainties $\sigma_{i}^{\rm data}$ were used 
in the denominator.

All the curves shown in this section display the exact point-by-point 
model prediction. In order to compare the model to the data, it is
 necessary to integrate over the bin size to obtain an average value 
equivalent to that  for  the data. In other words, the models were 
histogrammed into bins corresponding to the same bin sizes as the data.   
Each experimental point is counted as a degree of freedom and the 
comparison yields the results, listed in Table~\ref{chisq}.
\begin{table}
\begin{ruledtabular}
\caption{\label{chisq} $\chi^2$ per degree of freedom   
comparison between the data and  the four theoretical models. }

\begin{tabular}{ c c c } 

{\sl Model}     & $A_{t}$ (ndf=102) & $A_{et}$ (ndf=65)  \\ \hline
MAID2000        & 1.8            & 1.1  \\ 
SL              & 1.1            & 1.2  \\ 
DM              & 4.1            & 1.7\\
DMT             & 2.0            & 0.9 \\       
\end{tabular}
\end{ruledtabular}
\end{table}

The results of the $\chi^2$ comparison for the MAID, SL, and DMT
models give  very similar fits for the double spin asymmetry 
$A_{et}$. The differences  in the total $\chi^2$ are primarily 
determined by the comparison  with the single spin asymmetry $A_{t}$. 
On one hand the double spin asymmetry is characterized by 
the  $|M_{1+}|^2$ term,  which all the models describe reasonably well.
The target asymmetry on the other hand involves the 
imaginary part of interference terms and therefore depends on multipoles 
such as $E_{0+}$, $S_{0+}$, $M_{1+}$ and $S_{1-}$, which have larger 
uncertainties in the models.
In this respect the SL model considers all the second order processes, 
whereas MAID makes approximations for these terms. 
A dynamic approach of DMT accounts for these second order 
processes, but appears to give a similar fit as the MAID model.
The effective Lagrangian model of DM model does not include tails 
from higher resonances,  limiting the background description even further,
and may explain the large discrepancy with the polarization data.

        The DMT and MAID models were also observed to give similar
fits to each other for electron single spin $A_e$ observed
at lower $Q^2$ at JLab~\cite{joo} and  Mainz~\cite{bartch},
although both are in somewhat disagreement 
with those data.

\section{Summary}

Target and double spin asymmetries for the \delt decaying into 
$p$ and $\pi^0$ were extracted as a function of the pion center-of-mass 
angles, $\theta^*$ and $\phi^*$, and the momentum transfer $Q^2$. 
Comparison with some of the existing 
theoretical approaches was performed and  sensitivity to the different 
models was observed. A $\chi^2$ comparison shows (see table~\ref{chisq}) 
that the model with the best agreement with the data is the dynamical 
model of SL. The isobar model MAID and dynamic
models of DMT exhibited comparable fits in  reasonable agreement 
with the data.  Aside from the speculations about the various model 
sensitivities given here, a discussion of the technical differences 
which give rise to the differences in theoretical approaches
is beyond the scope of this article. Rather, it is the intent of this 
work to make available the unique experimental observables as constraints 
on all the models mentioned in the introduction.

\appendix*
\section{Multipole notation}
\label{app}

The cross section for electro-production in equations~(\ref{forsigma}) can also be written as a combination of Legendre Polynomials and their first and second derivatives. The coefficients of this expansion are the multipoles: $E_{l\pm}$, $M_{l\pm}$, and $S_{l\pm}$ \cite{Chew:1957tf}. The multipoles characterize the excitation mechanism (electric ($E$), magnetic ($M$), and coulomb or scalar($S$) type of photon) and the angular momentum of the final state $\pi N$. $l\pm$ refers to a state with a $\pi N$ relative angular momentum $l$ and total angular momentum $J=l\pm \frac{1}{2}$.

\begin{acknowledgments}
We thank R.M. Davidson, T.S. Lee and L. Tiator for the valuable help. 

We acknowledge the efforts of the staff  of the
Accelerator and Physics Division at Jefferson Lab
for making this experiment possible. This work was supported
by the U. S. Department of Energy, The U. S. National Science Foundation,
The French  Commissariat a l'Energie Atomic, the Italian Istituto 
Nazionale di Fisica Nucleare.
\end{acknowledgments}

\bibliography{article_edit}

\end{document}